# COMPREHENSIVE JOINT MODELING OF FIRST-LINE THERAPEUTICS IN NON-SMALL CELL LUNG CANCER

Authors and Affiliations

Benjamin K Schneider[1], Sebastien Benzekry[1,2*] and Jonathan P Mochel[1*]

[1]Iowa State University College of Veterinary Medicine, Ames, IA, U.S.A, [2]Team MONC, Inria Bordeaux Sud-Ouest, Institut de Mathématiques de Bordeaux, France. [*]: co-last authors.

## Abstract

First-line antiproliferatives for non-small cell lung cancer (NSCLC) have a relatively high failure rate due to high intrinsic resistance rates and acquired resistance rates to therapy. 57% patients are diagnosed in late-stage disease due to the tendency of early-stage NSCLC to be asymptomatic. For patients first diagnosed with metastatic disease the 5-year survival rate is approximately 5%. To help accelerate the development of novel therapeutics and computer-based tools for optimizing individual therapy, we have collated data from 11 different clinical trials in NSCLC and developed a semi-mechanistic, clinical model of NSCLC growth and pharmacodynamics relative to the various therapeutics represented in the study. In this study, we have produced extremely precise estimates of clinical parameters fundamental to cancer modeling such as the rate of acquired resistance to various pharmaceuticals, the relationship between drug concentration and rate of cancer cell death, as well as the fine temporal dynamics of anti-VEGF therapy. In the simulation sets documented in this study, we have used the model to make meaningful descriptions of efficacy gain in making bevacizumab-antiproliferative combination therapy sequential, over a series of days, rather than concurrent.

**Reference Nomenclature and Abbreviations for Equations**

afolate    The antifolate action of a cytotoxic drug i.e. pemetrexed

apo    Apomab aka DAB4, its chimeric derivative chDAB4, or PRO95780

auc    The area under the curve of drug concentration over time, usually interpreted as a measure of exposure

bev    Bevacizumab

car    Carboplatin

cc    Concentration, usually concentration as a function of time

cis    Cisplatin

doc    Docetaxel

dr5    The mechanism of action of Apomab, a monoclonal antibody against death receptor 5

eff    Effect

egfr    The egfr-based (endothelial growth factor receptor) mechanism of erlotinib

erl    Erlotinib

gem    Gemcitabine

kk    Rate of passage between compartments

microt    The microtubule-inhibition mechanisms of paclitaxel and docetaxel

n    Sum of primary all tumor volumes i.e. $v + v_i + z_1 + z_2 + z_3$

dnasub    The mechanism of action of gemcitabine whereby it masquerades as a functional nucleoside only to cause masked chain termination when incorporated into DNA

pac    Paclitaxel

| | |
|---|---|
| pem | Pemetrexed |
| plat | The class of cytotoxics whose mechanism involves platinum i.e. cisplatin, carboplatin, etc. |
| $Q_\alpha$, $Q_\rho$ | The effects whereby certain drugs limit the rate of growth of the tumor e.g. cell cycle arrest, nutrient supply disruption, etc. |
| $Q\gamma$ | The effect whereby certain drugs kill tumor cells directly |
| t | Time in days |
| v | Primary tumor volume in $cm^3$ |
| $v_i$ | Volume of tumor cells which are injured by vegf and microt drugs |
| vegf | The anti-vegf (vascular endothelial growth factor) mechanism of bevacizumab |
| $z_{1-3}$ | Irreversible cell death compartments |
| $w_d$ | A scaling factor between drug concentration and cytotoxic effect |
| $\lambda_d$ | Scaling factor between drug exposure and resistance |

## Introduction

With an estimated 135,000 deaths per year in the United States, lung cancer has the highest mortality rate of any cancer type (1). Approximately 85% of those deaths are attributable to non-small cell lung cancer (NSCLC). Often, NSCLC is not detected until late-stage disease as the early progression produces relatively few symptoms. First-line therapeutics for the management of metastatic or recurrent NSCLC (i.e. late-stage NSCLC) include combination chemotherapies (including platinum-based doublets) alone, or associated with immune checkpoint inhibitors (i.e. PD-1/PD-L1 or CTLA4 inhibitors) or antiangiogenics (e.g. bevacizumab, BEV). However, first- and second-line therapies have a relatively high failure rate. As an example, in the recent ARIES observational cohort study of first-line treatment involving bevacizumab the failure rate was approximately 51%.

Acquired or intrinsic drug resistance is a major cause for therapeutic failure in NSCLC. In a previous study of resected NSCLC by d'Amato et al., intermediate to extreme resistance to carboplatin was found in 68% of NSCLC cultures (vs. 63% and 40% for cisplatin and paclitaxel, respectively). Likewise, in the KEYNOTE-001 study completed last fall, NSCLC patients receiving pembrolizumab had an objective response rate of 19.4%, indicating that a vast majority of individuals did not significantly respond to therapy. Patients are also sometimes forced to cycle off chemotherapy due to excessive side-effects. Chemotherapeutics are notorious for broad and severe off-target effects that exacerbate underlying vulnerabilities – e.g. chronic kidney disease. After initial therapeutic options fail, many patients choose to enroll in clinical trials to receive experimental therapies (2). Though the prognosis for NSCLC patients is improving, at this time the 5-year survival rate in non-small cell lung cancer (NSCLC) is only approximately 20%. For the 57% of lung cancer patients first diagnosed with metastatic disease the 5-year survival rate is approximately 5% (3). Therefore, there is a critical need to improve the efficacy of existing therapeutics, to develop a better understanding of individualized NSCLC therapy, and to accelerate experimental therapeutic development. Mathematical modeling is a key strategy for addressing these needs, as its use in cancer is a proven method for solving problems in optimization and precision medicine (4–7).

Mathematical modeling of drug pharmacokinetics and pharmacodynamics is an extremely efficient method for optimization of therapeutic dosing schedules, without the considerable time and resource investment required to conduct a suite of in vivo clinical trials. Using this computational platform, one can leverage data from multiple studies, involving diverse patient populations, varying drug combinations and administration schedules, to build a series of "what if" scenarios and derive the best scheduling and dosing of therapeutic drug intervention in a given set of patients.

Bevacizumab is an important targeted therapy in treating NSCLC that has performed strongly when used in combination with a broad spectrum of antiproliferatives. Bevacizumab is an anti-angiogenic, and therefore inhibits the development of the neovascular architecture supporting tumor proliferation. Neovascular growth disruption is directly cytotoxic. Bevacizumab has also been shown to induce a transient period of perfusion normalization. This perfusion normalization improves the delivery of cytotoxics to the tumor resulting in improved efficacy of the cytotoxic. Because efficacy is improved, prescribing cytotoxics in combination with bevacizumab often allows clinicians to reduce the dosage of primary cytotoxics, thereby reducing the side-effect burden on the patient (8).

Previously, we have shown experimentally and described mathematically the time-course of this transient perfusion-normalization in a xenograft-murine model of NSCLC (9). In that experiment, pemetrexed-cisplatin (PEM/CIS) was administered sequentially with bevacizumab (BEV). By varying the gap between bevacizumab and cytotoxic administration, we were able to estimate the time-course of vascular normalization, as well as several other parameters used to build a pharmacokinetic-pharmacodynamic (PKPD) description of NSCLC growth and response to BEV-PEM/CIS.

In our previous model description, we could predict that administering BEV and PEM/CIS sequentially with a gap of 1 day, rather than concurrently, would improve the efficacy of this combination (quantified as final tumor volume) by more than 50% without the need for increasing therapeutic doses. However, optimal scheduling of this combination in humans has yet to be verified with large sets of clinical data, and the model needs to be generalized to other combination therapies, including immune checkpoint inhibitors.

Our previous preclinical study is a demonstration of a proven paradigm of therapeutic development that could become especially impactful in NSCLC. In this paradigm, scientists develop novel and individualized therapeutic strategies from existing and approved therapeutics (2,10). Additionally, there is a wealth of data currently available from clinical data-sharing services such as Vivli, CSDR, Project Datasphere, YODA and others. When using previously produced data, researchers obviate the need for human or animal participants. The cost and time required to clean and organize the data, build the model, evaluate the model, and simulate clinical trials is drastically less than that required when designing and performing new clinical trials. Clinical trials are usually only designed to test a sparse number of medications and scheduling strategies over the course of the study. Via simulation, we can trial a vast number of medication combinations and scheduling strategies simultaneously.

Data from clinical trials will not always be directly comparable due to, for example, variations in experimental design, differences in operating procedure, and differences in patient cohort characteristics. This variation can be an advantage of pooling datasets from separate trials if the statistical framework for analysis is able to account for and elucidate any potential sources of variation. Non-linear mixed effects (NLME) modeling of tumor proliferation and response is a proven framework for pooling data to model-build (9,11,12)

In this study, we have implemented these principles to build a unified, semi-mechanistic model and platform for scheduling bevacizumab in combination with several approved therapeutics in NSCLC. Specifically, we have collated individual tumor progression data from 11 different clinical trials (> 8000 stage II through stage IV, metastatic, and non-metastatic patients) involving BEV and multiple chemotherapies such as PEM, CIS, apomab, paclitaxel, carboplatin, gemcitabine, and erlotinib. These data were made available to us through a data sharing agreement with vivli.org. Our overall objectives in this study are to (1) generalize our model of NSCLC growth and response to BEV-PEM/CIS to the greater set of combination therapies and modes of action represented in our large clinical database, and (2) characterize the time-course of resistance for those registered therapeutics in humans.

## Methods

### Literature Search and Review

To build our model, we sought access to data using very broad criteria. Briefly, we were interested in requesting access to data from clinical trials where bevacizumab had been used in combination with other therapeutics to treat NSCLC. To build on previous modeling efforts, it was necessary for most studies to include records of individual patient tumor sizes over time. After reviewing several potential data access providers with which we could partner, we applied for data access through the Clinical Study Data Request (CSDR) portal. We were able to identify 11 different studies available through CSDR's platform which met our requirements (Table 1) and were permitted access to a secure server containing the datasets beginning on January 6th, 2020.

Due to contractual obligations between the data owners (Roche, Eli Lilly) and data access provider (CSDR), our project and access to the datasets were moved to a secure server managed by Vivli on September 22nd, 2020. During this transfer process, we requested access to, and were granted access to, 5 additional datasets involving immune-checkpoint inhibitors and NSCLC which were not available at the time we first applied for access through CSDR.

**Data Processing and Collation**

Data annotation, units, organization, and shorthand is highly variable between datasets. This variability makes collating and preparing large datasets for mathematical modeling a challenging, and highly error-prone task.

To reduce the possibility of introducing mistakes into the datasets during collation, we systematized the iterative collation of datasets. First, we identified the comma separated value files containing the most relevant base information. The information we were interested in for initial model-building was individual, tumor identifier, tumor type, tumor size, and drug administration details all organized longitudinally. Data were then imported into R (version 3.5.2) to be normalized for import into our NLME parameter estimation software (Monolix Suite version 2020R1, Lixoft). Normalization consisted of formatting the data as recommend by Lixoft (15), as well as matching units to an internal dictionary of standards – tumor measurements in cm, amount administered in mg, etc.

After each dataset was produced, the data was explored visually with a combination of R as well as Datxplore (Monolix Suite version 2020R1) to check for inconsistencies. Finally, the individual datasets produced in analyzing each individual study were bound into one single comma separated value file. As a final quality check, the final dataset was re-imported in R, and subset down to individual studies. Then, the processed studies were compared with the raw files from the clinical trials for consistency. Data were both received and maintained in a fully anonymized format to protect patient privacy.

**Non-Linear Mixed Effects Modeling and Characterizing Individual Variation**

The recorded data ($y_{ij}$) were pooled and used to estimate model parameters via the stochastic approximation expectation maximization algorithm (SAEM) as implemented in Monolix. After estimating population parameters ($\boldsymbol{\mu}$) and variance, individual parameters ($\boldsymbol{\phi}_i$) were estimated using the modes of the individual estimated posterior distributions. The posterior distributions were estimated using a Markov-Chain Monte-Carlo (MCMC) procedure. NLME models were written as previously described (Pelligand et al., 2016; Sheiner & Ludden, 1992) (Equation 1).

**Equation 1:**

$$y_{ij} = F\big(\boldsymbol{\phi}_i,\ \boldsymbol{\beta}_i,\ t_{ij}\big) + G\big(\boldsymbol{\phi}_i,\ t_{ij}\big) \cdot \varepsilon_{ij} \quad | \quad \boldsymbol{\varepsilon}_{ij} \sim \mathcal{N}\big(\mathbf{0},\ \sigma^2\big)$$

$$\boldsymbol{\phi}_i = h(\boldsymbol{\mu},\ \boldsymbol{\eta}_i,\ \boldsymbol{\beta}_i) \quad | \quad \boldsymbol{\eta}_{ij} \sim \mathcal{N}(\mathbf{0},\ \boldsymbol{\Omega},\ \omega^2)$$

$$j \in 1, \dots, n_i, \qquad i \in 1, \dots, N$$

Model predictions ($F(\boldsymbol{\phi}_i,\ \boldsymbol{\beta}_i,\ t_{ij})$) for the $i$th individual at the $j$th timepoint were written as a function of individual parameters, individual covariates ($\beta_i$), and time ($t_{ij}$). The residuals were modeled as $G(\boldsymbol{\phi}_i,\ t_{ij}) \cdot \boldsymbol{\varepsilon}_{ij}$.

Individual parameters are modeled with function $h(\boldsymbol{\mu},\ \boldsymbol{\eta}_i, \boldsymbol{\beta}_i)$. Interindividual variability, $\boldsymbol{\eta}_i$, are distributed normally with mean $\mathbf{0}$, variance-covariance matrix $\boldsymbol{\Omega}$, and variance $\omega^2$. Typically, $h(\boldsymbol{\mu},\ \boldsymbol{\eta}_i,\ \boldsymbol{\beta}_i)$ is a lognormal link function (Equation 2). In cases where $\phi_i$ is bounded, $h(\boldsymbol{\mu},\ \boldsymbol{\eta}_i,\ \boldsymbol{\beta}_i)$ was typically a lognormal link function (Equation 3).

**Equation 2:**

$$\phi_i = \mu \cdot e^{\eta_i + \beta_i}$$

**Equation 3:**

$$log\left(\frac{\phi_i}{1 - \phi_i}\right) = log\left(\frac{\mu}{1 - \mu}\right) + \eta_i + \beta_i$$

**Model Building**

PKPD models were built in multiple development phases. First, we reproduced previously established models of therapeutic PK within our modeling project using Mlxtran. Then, we created several sample sets with between 5% and 10% of the complete dataset for initial model building. This shortened calculation time and reduced computational complexity. Next, we identified our preferred base candidate models for PD and implemented them in Mlxtran. Finally, using manual exploration and cursory SAEM parameterizations, we were able to determine reasonable parameter estimates with which to initialize our parameter search.

The PK portion of our PKPD model was implemented using previously published human PK models and parameter values. PK models were written in Mlxtran. We verified PK model integrity by simulating trivial scenarios with known outcomes using Mlxplore. If outcomes simulated deviated from expectations, the code was reviewed for inconsistencies.

To begin building our model of NSCLC PKPD, we used exploration to parameterize two base models of NSCLC PD. The first model parameterized was the traditionally used Claret model of tumor growth and response to anti-proliferatives. The second base model implemented was built on our previous Gompertzian model of BEV-PEM/CIS published elsewhere.

To initially parameterize these candidate models, we visually explored potential initial parameter spaces using smaller sample subsets. Once we had a set of potential initial parameter values, we ran the SAEM algorithm on these sample sets to determine numerical stability of these parameter values. This process would result in loops of exploration and parameterization leading to further exploration and parameterization. Once we had implemented a PKPD model in Mlxtran and had reasonable initial parameter estimates, we were able to fit the model to the full dataset using the SAEM method. After fitting base models of PKPD, we were able to evaluate and iteratively improve on the deficiencies in fit using numerical experimentation, model exploration, goodness-of-fit metrics, traditional analysis, as well as reviewing previously published theories of NSCLC response to various therapeutics. For example, using simulation engines like Simulx and Mlxplore, we were able to test whether known biological phenomena were reliably reproduced by the model. Using model evaluation tools (below), we were able to test whether our fit appropriately met the assumptions made for NLME modeling. If the model did not produce individual fits which closely matched measurements, we used fundamental compartmental modeling concepts to add parameters, and therefore flexibility, to the model.

Our primary structural concerns were modeling individual tumor growth and response, modeling both acquired and intrinsic resistance, modeling individual drug effects and interactions, as well as modeling the timescale of perfusion enhancement via bevacizumab. Modeling perfusion enhancement via bevacizumab was especially important as it was the primary structure that would allow us to determine the optimal predicted scheduling of sequential bevacizumab and various anti-proliferatives.

**Model Evaluation**

We employed state-of-the-art model building techniques to develop our model into a meaningfully complete description of the variation in the dataset and evaluate the quality of model. This broadly required us to evaluate quality of fit, validate assumptions made about variance, test correlation between individual parameters with both other individual parameters as well as covariates, consider various statistical formulations of individual parameters, determine precision of individual parameter estimates, and finally measure models against one another.

Quality of fit was determined using both goodness-of-fit plots and summary statistics. Stability of parameter estimates was assessed by both inspection of SAEM search, attainment of auto-stop criteria as implemented in Monolix, and whether randomized (but still local) initial starting parameters converge to the same set of parameter estimates. Accuracy of individual fits was assessed using a sample of individual plots, an observations-vs-predictions plot using the full conditional distribution, a scatter plot of the residuals, as well as the corrected Bayesian information criteria (BIC) – estimated via importance sampling).

Assumptions of variance were validated by plotting the conditional distribution against theoretical distributions as well as standard statistical tests. Within the NLME framework, random effects and residuals are assumed to be predictably distributed – usually normally or functionally-linked to normal. We used the Shapiro-Wilk test of normality to determine normality, and Van Der Waerden test to determine symmetry of distributions about 0.

Correlation between individual parameters and other individual parameters or covariates were tested with a Pearson's correlation test and ANOVA, respectively. Plotting these relationships assisted in determining the nature of these correlations.

Precision and accuracy of the final model was assessed to evaluate models against one another. Precision of parameter estimates was made using relative standard error (RSE). Overall model quality of fit was evaluated BICc. Diagnostic plots assisted in comparing models with similar overall performance.

**Clinical Trial Simulations**

In our clinical trial simulations, we hoped to determine what benefit (if any) sequentially administering therapy with bevacizumab produced. To do this, we took the individual fits from the experimental arms of the trials and simulated them with various gaps between bevacizumab and treatment using Simulx (Monolix Suite version 2020R1, Lixoft). The highest performing gap was then compared with the experimental standard.

**Study Details**

*Randomized, Open-Label, Phase 3 Study of Pemetrexed Plus Carboplatin and Bevacizumab Followed by Maintenance Pemetrexed and Bevacizumab Versus Paclitaxel Plus Carboplatin and Bevacizumab Followed by Maintenance Bevacizumab in Patients With Stage IIIB or IV Nonsquamous Non-Small Cell Lung Cancer*

Between 2008 and 2014, 939 patients were administered either bevacizumab + pemetrexed + carboplatin (experimental arm) or bevacizumab + paclitaxel + carboplatin (comparator arm) to treat phase 3 NSCLC. Primary outcome was overall survival time from baseline to date of death (any cause). ClinicalTrials.gov Identifier: NCT00762034

*A Study of Pemetrexed and Bevacizumab for Participants With Advanced Non-Small Cell Cancer*

Between 2009 and 2013, 109 patients were administered bevacizumab + pemetrexed + carboplatin (single treatment arm) to treat stage IIIB or stage IV NSCLC that was not amenable to curative therapy. Primary outcome was progression free survival from baseline. Progression was scored using response evaluation criteria in solid tumors (RECIST criteria). ClinicalTrials.gov Identifier: NCT01004250

### *A Study of PRO95780 in Patients With Previously Untreated, Advanced-Stage Non-Small Cell Lung Cancer (APM4074g)*

Between 2007 and 2010, 128 patients were administered either bevacizumab + carboplatin + paclitaxel + PRO95780 (experimental arm), or bevacizumab-pemetrexed/carboplatin (single treatment arm) to treat stage IIIB or stage IV NSCLC that was not amenable to curative therapy. Primary outcome was progression free survival from baseline. Progression was scored using response evaluation criteria in solid tumors (RECIST criteria). ClinicalTrials.gov Identifier: NCT00480831

### *A Study of Avastin (Bevacizumab) in Patients With Non-Squamous Non-Small Cell Lung Cancer (NSCLC)*

Over a period of 6 years starting in 2005, 1044 patients with advanced or recurrent non-squamous NSCLC were randomized to any of three experimental arms. The first two arms were cisplatin + gemcitabine + bevacizumab combination therapy with bevacizumab administered at either a high dosage or low dosage. The third arm was a placebo comparator where patients were administered cisplatin + gemcitabine + placebo. The primary outcome measured was progression-free survival. Efficacy and safety were tracked as secondary outcomes. Efficacy was defined as a combination of survival duration, time to treatment failure, response rate, and duration of response.

Safety was measured through a state-of-the art panel of laboratory values.

ClinicalTrials.gov Identifier: NCT00806923

   *A Study of Avastin (Bevacizumab) in Combination With Carboplatin-Based*

*Chemotherapy in Patients With Advanced or Recurrent Non-Squamous Non-Small Cell*

*Lung Cancer*

   From 2008 to 2012, 303 patients were administered a combination of bevacizumab and carboplatin with either 7.5 mg/kg bevacizumab or 15.0 mg/kg bevacizumab. The treatments were being used to treat patients with advanced or recurrent non-squamous NSCLC. The primary outcome measured was the percentage of patients with either a complete response or partial response. Progression-free survival and duration of response were the secondary outcomes measured.

ClinicalTrials.gov Identifier: NCT00700180

   *A Study of Bevacizumab in Combination With First- or Second-Line Therapy in*

*Subjects With Treated Brain Metastases Due to Non-Squamous NSCLC (PASSPORT)*

   Starting in 2006, 115 participants with metastatic non-squamous NSCLC with

previously treated central nervous system metastases were administered an

experimental combination of bevacizumab with either first-line or second-line

chemotherapy. The trial lasted for 3 years. The primary outcome measured was the

percentage of participants with symptomatic NCI CTCAE (16) Grade ≥ 2 CNS

hemorrhage. Secondary outcomes measured included patient overall survival and

adverse effects. ClinicalTrials.gov Identifier: NCT00312728

   *A Study of Avastin (Bevacizumab) in Patients With Non-Squamous Non-Small*

*Cell Lung Cancer With Asymptomatic Untreated Brain Metastasis*

Between 2009 and 2012, 91 patients with stage IV NSCLC were sorted into one of two experimental groups. The first group received bevacizumab + carboplatin + paclitaxel combination therapy and the second group received bevacizumab + erlotinib combination therapy. The primary outcomes measured were progression-free survival at 6 months, percentage of participants with disease progression, and time to disease progression or death. Secondary outcomes included percentage of patients achieving complete response and percentage of patients achieving partial response. ClinicalTrials.gov Identifier: NCT00800202

*Effects of Two Doses of rhuMAB-VEGF Antibody in Combination w/Chemotherapy in Subjects With Locally Advanced or Metastatic Lung Cancer*

The details and results from this clinical trial were published in 2004 [https://doi.org/10.1200/JCO.2004.11.022]. In this phase II trial, 99 patients were randomly assigned either to bevacizumab + carboplatin + paclitaxel (experimental arm) or carboplatin + paclitaxel alone (comparator arm). The primary efficacy measures were time to disease progression and best confirmed response rate. ClinicalTrials.gov Identifier: N/A

*Safety and Efficacy Study of Avastin in Locally Advanced Metastatic or Recurrent Non-small Lung Cancer (NSLC) Participants*

996 participants were enrolled in this study between its start and finish, 2007 and 2013. Patients were administered several cycles of bevacizumab before beginning chemotherapy. After chemotherapy, patients were cycled back to bevacizumab as a maintenance therapy. The primary outcome measurement was the percentage of patients with adverse effects. Among the secondary outcomes measured were the number of cycles of therapy tolerated, percentage of patients with complete or partial response, eastern cooperative group (ECOG) performance status grades, and progression free survival. ClinicalTrials.gov Identifier: NCT02596958

    *A Study of Bevacizumab Versus Placebo in Combination With Carboplatin/Paclitaxel in Participants With Advanced or Recurrent Non-Squamous Non-Small Cell Lung Cancer Who Have Not Received Previous Chemotherapy*

    This randomized, double-blind, placebo-controlled study was conducted between 2011 and 2017. 276 participants were distributed to either a group receiving bevacizumab + carboplatin + paclitaxel (experimental arm) or carboplatin + paclitaxel + placebo (active comparator arm). Progression free survival was the primary outcome measured, and several survival and lab-value related secondary outcomes were also measured. ClinicalTrials.gov Identifier: NCT01364012

    *A Study of Avastin in Combination With Chemotherapy for Treatment of Colorectal Cancer and Non-Small Cell Lung Cancer (ARIES)*

    In the ARIES study, approximately 4,000 patients with locally advanced or metastatic NSCLC (or a similar colorectal cancer – CRC), who were also receiving a combination bevacizumab therapy, were observed between 2006 and 2012 for disease progression. Several data-points were collected over the period to measure safety and efficacy of bevacizumab for NSCLC or CRC. The collected data include progression-free survival and physical tumor biopsies. This study was largely ignored for semi-mechanistic modeling purposes as it included few tumor measurements. ClinicalTrials.gov Identifier: NCT00388206

**Results**

**Data Summary**

Data were received in directories of fully anonymized dataframes (e.g. excel files, SAS files, .csvs) organized by general category of data. For initial data processing, data were explored for bulk trends, standardized, and unsuitable data were removed. Studies were then collated in a single dataframe designed for use with the Monolix suite. Between the 11 studies, 3686 patients' data were determined to be potentially suitable for analysis. After exploring the datasets in greater detail, we determined that only 2586 patients (between the studies) had all the data necessary to create models of tumor growth and response – unique patient IDs, a time recorded for each dosing and measurement event, tumor measurements, and therapeutic administration details for each patient. Data that were unsuitable for analysis were either missing proper documentation detailing the contents of the dataframe – e.g. dictionary for column ids – or were missing data necessary for longitudinal modeling – e.g. unique patient ids or dosing events.

We chose to model individual tumor longest diameter time-course as our independent variable, as the sum of the longest diameter (SLD) typically do not perform as well as individual tumor diameter in semi-mechanistic models of tumor growth, and individual tumor longest diameter is truer to the biology of the disease {cite}. We were not able to separate inter-individual variability (IIV) and inter-occasion variability (IOV) resulting from patients with more than 1 tumor being measured. Statistically, all patient-tumor combinations were treated as unique subject-occasions. This is a biological oversimplification, as multiple tumors within a single individual most likely have related individual parameters, but this compromise provided a good balance between modeling systems biology and model identifiability. Among the subject-occasions, we had data from 6197 unique tumors belonging to 2586 patients.

After a short period of testing, we imposed several further restrictions on the dataset to better facilitate modeling. The first condition was that tumors were required to have been measured 3 or more times to qualify for inclusion. This reduced the number of unique tumor ids to 4701 and unique individuals to 2036. Then we removed monotonic non-responders from the dataset. If for each sample $y_{i,t}$, for individual $i$ at time $t$, $y_{i,t}$ was greater than or equal to $y_{i,t-1}$ we labeled the tumor as a monotonic non-responder and excluded it from the dataset. This reduced the number of individual tumors to 4473 and individual ids to 1977. After removing these data, we were left with 4450 tumors and 1963 individuals. These restrictions imposed on the initial dataset representing 2586 patients reduced the number of samples from 29885 to 26515. This is an approximate 11% reduction in data. Lastly, because of our limited CPU resources, we chose to only work with approximately 5% of the data from each study (randomly allocated by subject-tumor pairs). The final dataset used for model building detailed tumor growth time-course for 250 individual tumors from 221 patients.

We also imposed an artificial condition on the lower limit of quantification to reduce the chance that noisy measurements would produce numerical instabilities in the SAEM search. As an example, in some of the datasets, clinicians would record size = 0 if they could not find a tumor. If during the next visit, the clinician would find the tumor again, the clinician would log a size greater than 0 for the tumor. Records like these would occasionally cause convergence issues with the SAEM algorithm. To remedy this, we set the lower limit of quantification as 1 mm in diameter.

**PKPD Model Building**

Individual PK parameters and models were estimated in few of the clinical trials used in this study. Therefore, population PK models were collated from various publications involving the relevant therapeutics, and IIV was not included in the final PK model as it led to structural unidentifiability (Table 5.1).

As a first step, we wrote our previous model of bevacizumab-pemetrexed/cisplatin therapy in Mlxtran and validated the model against the dataset produced in various treatments receiving any combination of the medications. The units of the previous publication were relative to fluorescence, so results were not directly translatable. Also, our previous model only dealt with combination bevacizumab-pemetrexed/cisplatin and we used a simple scaling factor against concentration to evaluate effect. This prevented us from being able to validate against other medications. We only simulated over short periods in our previous work which made the effects of acquired resistance relatively inconsequential, however this effect was readily available in various patients within the Vivli dataset. Taken together, this suggested that we first start by modifying the Gompertzian growth and cytotoxic kill effect equations to account for an expanded set of medications and new units, and later attempt to capture the dynamics of resistance.

To create initial parameterizations for our pharmacodynamic modeling, we subsetted to small samples of the full dataset to perform short experiments as well as visual exploration. Two primary models were evaluated for the basic description of tumor growth and response; (1) the Gompertzian model of tumor growth and (2) the Claret model of tumor growth (6,13). As both models are relative to tumor volumes, we modeled the tumors as spherical – an oversimplification as tumors are often oblong.

Using the Bayesian information criteria as a parsimonious method of cross-evaluating models, we found that the Gompertzian model outperformed the Claret model of tumor growth at several layers of structural model-building. In the Gompertz model of tumor growth (Equation 4), the unperturbed tumor grows at rate $\alpha$ and is exponentially limited in growth by parameter $\beta$. $v_c$ is a scaling factor relating individual tumor cell turnover to volume. It was set to $10^6$ cells/mm$^3$ which is the classical assumption of approximate number of cells per unit volume (14).

## Equation 4

### (4a)

$$\frac{dv}{dt} = \left( \alpha \cdot Q_\alpha - \beta \cdot log\left(\frac{v}{v_c}\right) \right) \cdot v - Q_\Upsilon \cdot v$$

$$log(Q_\alpha) = -\left( 1 + w_{bev_\alpha} \cdot cc_{bev}(t - \tau) \right) \cdot (eff_{microt} + eff_{vegf})$$

$$Q_\Upsilon = \left( 1 + w_{bev_\Upsilon} \cdot cc_{bev}(t - \tau) \right) \cdot \left( eff_{plat} + eff_{afolate} + eff_{dr5} + eff_{egfr} + eff_{dnasub} \right)$$

### (4b)

$$\frac{dv}{dt} = \left( \alpha - \beta \cdot log\left(\frac{v}{v_c}\right) \right) \cdot v - Q_\Upsilon \cdot v - Q_\rho \cdot v + kk_i \cdot v_i \cdot p$$

$$\frac{dv_i}{dt} = Q_\rho \cdot v + kk_i \cdot v_i \cdot p - Q_\Upsilon \cdot v_i$$

$$Q_\delta = \left( 1 + w_{bev_\delta} \cdot cc_{bev}(t - \tau) \right) \cdot (eff_{microt} + eff_{vegf})$$

$$Q_\Upsilon = \left( 1 + w_{bev_\Upsilon} \cdot cc_{bev}(t - \tau) \right) \cdot \left( eff_{plat} + eff_{afolate} + eff_{dr5} + eff_{egfr} + eff_{dnasub} \right)$$

In Equation 4a, $Q_\alpha$ and $Q_\Upsilon$ are the antiproliferative effects resulting in growth reduction and irreversible cell death, respectively. Chemotherapeutics which acted on the microtubules – docetaxel and paclitaxel – along with the direct effect of bevacizumab were included in $Q_a$. All other drugs were treated as drugs resulting in irreversible cell death. Transient enhancement in efficacy via perfusion normalization by bevacizumab was modeled as occurring at time $(t - \tau)$ to account for the time delay between administration and efficacy enhancement i.e. $\tau$. We found Equation 4a to heavily exaggerate the effect of bevacizumab in limiting cell growth rates. To account for this, we used a second compartment which represented reversible cell injury, $v_i$, from which cells could return from (Equation 4b). Return rate is governed by intercompartmental transfer rate $kk_i$ as well as proportion of repaired cells returned to the unperturbed cycle of proliferation, $p$. We found this better represented the growth limiting effects of bevacizumab, paclitaxel, and docetaxel without being as exaggerated of an effect as modeled in Equation 4a.

Irreversible cell death was modeled as occurring over a series of transitions between several compartments with intercompartmental transfer rate $kk$. The final tumor volume, a summation of the primary tumor volume and death compartments ($z_1$, $z_2$, and $z_3$,) as well as injured cell volume $v_i$, was then transformed to a tumor diameter to match the independent variable in the dataset (equation 5).

**Equation 5**

$$\frac{dz_1}{dt} = Q_\gamma \cdot v - kk \cdot z_1$$

$$\frac{dz_2}{dt} = kk \cdot z_1 - kk \cdot z_2$$

$$\frac{dz_3}{dt} = kk \cdot z_2 - kk \cdot z_3$$

$$n = v + z_1 + z_2 + z_3 + v_i$$

$$Tumor\ Diameter = 2 \cdot \left(\frac{3n}{4\pi}\right)^{1/3}$$

For the pharmacodynamic effect on the tumor growth, we started with a simple version of our previous model whereby each drug's concentration was scaled by a single parameter (represented by $w$ for weighting) and the sum of those concentrations determined cytotoxic effect (Equation 6a). That proved slightly unstable, so we eventually grouped the sum of those effects under an inverse logit function so that their sum would be limited. However, this also resulted in parameter instability as there was no upper cap on estimates (Equation 6b). After several more iterations of the model, we settled on a pharmacodynamic model that individualized drug effects relative to those that shared their mechanism of action. We also implemented a model of resistance. In this model, the cancer became increasingly more resistant (rate governed by parameter $\lambda$) to treatment as a function of exposure (AUC) – Equation 6c. A whole diagram of the model is available for review in Figure 5.1.

**Equation 6**

**(6a)**

$$eff_d = (w_d \cdot cc_d)$$

$$eff_{total} = \sum w_d \cdot cc_d$$

$$d \in cis, car, pem, apo, erl, gem, pac, doc, bev$$

**(6b)**

$$eff_d = (w_d \cdot cc_d)$$

$$eff_{total} = \gamma \cdot \left( invlogit \left( \sum w_d \cdot cc_d \right) - 0.5 \right)$$

$$d \in cis, car, pem, apo, erl, gem, pac, doc, bev$$

**(6c)**

$$AUC_d = \int cc_d(t)$$

$$\begin{bmatrix} single\ drug\ eff & \equiv & log(eff_m) = log(w_d \cdot cc_d) - (\lambda_d \cdot AUC_d) \\ paired\ drug\ eff & \equiv & log(eff_m) = log\left(w_{d_{1-rel-2}}(cc_{d_1} + w_{d_2}cc_{d_2})\right) - \lambda_{d_{1-rel-2}}(AUC_{d_1} + \lambda_{d2}AUC_{d2}) \end{bmatrix}$$

$$d \in cis, car, pem, apo, erl, gem, pac, doc, bev$$

$$m \in plat, afolate, dr5, egfr, dnasub, microt, vegf$$

Individual variability was modeled using the standard log-link formulation and initial tumor volumes were fixed to the measurement of the tumor at time 0, relative for each individual. The only except was parameter *p* which was fixed between 0 and 1 using a logit-link function (Equation 7). Measurement error was modeled using the equation titled *combined 1* in Monolix, i.e. a single additive term (a) added with a single proportional term (b).

### Equation 7

log-link: $\phi_i = \phi_{pop} \cdot e^{\eta_i}$

logit-link: $logit(\phi_i) = logit(\phi_{pop}) + \eta\_i$

$$v(t=0) = y(t=0)$$

## Model Diagnostics

Our model diagnostics suggest a stable and precise fit that largely fits the assumptions necessary to draw conclusions. We observed that the SAEM search was stable and reliable when estimating our final set of parameter estimates (Figure 5.2). We were not able to perform a full convergence assessment because of computational constraints, but our experiments in subsets of the data suggest that results are stable regardless of relative initial parameter estimates. Individual fits were reasonably descriptive with both a well described response and rebound after treatment cessation (Figure 5.3). After seeing evidence of correlations (via Pearson's test) between individual effects, we inspected those correlations using the full posterior plot of individual effects in parameter $\eta_L$ vs $\eta_K$ where $L$ is not equal to $K$. We found that though correlations existed, the slope of the correlations was nearly zero and they were likely just natural artifacts that appear when working with large datasets (increased statistical power). An even spread of observations vs. individual predictions suggests that our model has no major structural misspecifications and that our error model was well specified (Figure 5.4). However, formal tests for residual normality and centering on zero failed. This is likely because of our use of the initial measured tumor volume as a covariate (residual is equal to 0) and the artificial implementation of a universal lower limit of quantification (observations fixed to 0.1). Once removing the censored points and the points measured at time 0, our residual error model aligns much more closely with the theoretical model. Precision of parameter estimates is extremely high with low dependency between estimates. Full parameter estimates, IIV, and RSEs are reported in Table 2. The visual predictive check (VPC) was informative as to wholistic model performance. Although the clinical trials were not matched, the VPC still indicates overall high-quality fit (Figure 5.4). Spread of individual parameters meets the Kolmogorov-Smirnov test for normality.

**Clinical Trial Simulations**

In our clinical trial simulations, we found an unexpected result. When we simulated the exact conditions of the clinical trial, but with a gap that was varied between administering bevacizumab 5 days earlier than scheduled and 5 days later than scheduled (at steps of 1 days), we found almost no difference between treatment groups. Put another way, administering bevacizumab before cytotoxics did not increase efficacy as expected (Figure 5.5 and Table 5.3).

We wanted to further investigate this outcome, so we simulated a human analog of our 2018 trial in mice using the full posterior distribution as estimated in Monolix as a virtual population pool. We largely found the same result (Figure 5.6 and Table 5.4).

As one last test, we began simulating individual IDs and comparing the outcome at various gaps from administering bevacizumab 5 days earlier to administering bevacizumab 5 days later. Explored in this way, it became clear that administering bevacizumab before pemetrexed/cisplatin did produce a significant effect in some patients, but not in all (Figure 5.7).

## Discussion

Our primary goal for this project was building a comprehensive semi-mechanistic model of NSCLC growth and response to various clinical anti-proliferatives. To that end, we have largely been successful in developing a model to explain the variation we see in the data. Our model captures the antiproliferative effects of the 11 different therapeutics used across the clinical trials as well as intrinsic and acquired resistance. We have pooled the best available and published pharmacokinetic models of the therapeutics involved, and we have used population estimates for each individual patient. This compromise likely slightly inflates the estimates of pharmacodynamic variation. Guided by our previous findings, we were also able to capture the transient enhancement of perfusion resulting from anti-VEGF therapy (bevacizumab). Individual predictions are relatively precise and our model captures the well-described rebound in growth after treatment cessation in non-small cell lung cancer.

We have also built that model with several innovations from our previous efforts. We have used AUC as a measure of exposure to model acquired resistance. Intrinsic resistance has been folded into the distribution of $w_d$ (i.e. weighting) terms in our system of equations. We also found that having a second cytotoxic effect for reversible cellular injury allowed us to capture direct effects of bevacizumab as well as capture the effect of medications known to cause reversible cell injury (i.e. paclitaxel and docetaxel). Our largely modular form of differential equations also provides a natural avenue for adding more drugs to the model. We have also provided a robust set of parameter estimates and have made our model code freely available for future research in non-small cell lung cancer.

When evaluating our model, we found robust evidence to support our choice of model structure. Parameter estimates and individual predictions were made with relatively high precision. This is likely due to the largeness of the dataset included. Model structure was based on biological mechanisms making interpretation of parameters relatively natural – e.g. $\lambda$ parameters define the rate of acquired resistance vs. exposure – and lending the model the longevity afforded by mechanistic modeling. On parameter estimate interpretation, we have used relatively simple naming heuristics to aid in interpretation. As stated above, $\lambda$ parameters define the rate of acquired resistance vs. exposure. $w_d$ parameters weight drug action against the tumor i.e. the larger the $w_d$ parameter, the larger the action the drug takes proportional to both the tumor size and concentration of the drug in plasma. The parameter titled imv_r_perc indicates the proportion of cells in the injured volume which return to unperturbed tumor growth. Unperturbed tumor growth is governed by parameters $\alpha$, exponential rate of growth of tumor, and β, exponential rate of decrease in growth rate due to nutrient supply limitations.

We acknowledge several weaknesses in our approach which must be addressed in future studies. Due to limited computing resources and poor documentation, we were unable to work with larger (approaching 95%) portions of the dataset. The model was not validated against data not used in training the fit. In the future, we will use individualized Bayesian predictions based on initial measurements of tumor growth and response to test if our model can make accurate predictions of the observed late stages of response. We also have a large amount of data on cutting-edge first-line therapies i.e. pembrolizumab and other immune-checkpoint inhibitors. This data was granted to us through a sub-request, and has not been able to be included in this modeling project. There are also several covariates which have not been included or tested in the model. To make individual predictions meaningful, they must be matched against patient characteristics and lab values.

One of the primary features we hoped to capture with our model was the transient enhancement of drug delivery after bevacizumab administration. Theoretically, this transient enhancement would drive the synergism between bevacizumab and other antiproliferatives. A natural conclusion, and a conclusion supported by previously published clinical papers, is that administering bevacizumab before other antiproliferatives should result in the greatest reduction in tumor size. Unexpectedly, we found through simulation that this result is only true in some cases. Moreover, in some individuals, delaying the bevacizumab until after other therapeutics provided the greatest reduction in tumor volume. Why this is so is not readily suggested by the model developed in this study. To solve this problem, our research group will simulate more clinical trials to determine if there is parameter clustering or some covariate which might predict what the optimal gap in any given patient might be.

Making individual predictions is the long-term goal for this modeling project. This first phase of research was meant to establish a robust preliminary model structure which explained a great deal of variation in the data. Expanding our dataset past 5% of the cleanest data, including covariates and individual patient characteristics, using simulation to find what might predict ultimate gap between bevacizumab and combination antiproliferatives, and refining the structure of our model will give us the full set of tools to develop tools to take individual patient data, and from that data individualize therapy.

**Tables and Figures**

Table 1 Pharmacokinetic Parameters

| Therapeutic | V1 | V2 | V3 | Q1 | Q2 | k12 | k21 | Cl | Source |
|---|---|---|---|---|---|---|---|---|---|
| Units | mL | mL | mL | mL/day | mL/day | day$^{-1}$ | day$^{-1}$ | mL/day | - |
| bevacizumab | 2804.12 | - | - | - | - | 0.223 | 0.215 | 216.291 | (17) |
| cisplatin | 22300 | 77000 | - | 456000 | - | - | - | 6480 | (18) |
| pemetrexed | 12900 | 3380 | - | 20736 | - | - | - | 131904 | (19) |
| apomab* | 3970 | 3840 | - | 793 | - | - | - | 328 | (20) |
| paclitaxel | 229000 | 856000 | 30300 | 3216000 | 5112000 | - | - | 10296000 | (21) |
| carboplatin | 11900 | 8230 | - | 2172000 | - | - | - | 177120 | (22) |
| gemcitabine | 15000 | 15000 | - | 1008000 | - | - | - | 3888000 | (23) |
| docetaxel | 7900 | - | - | - | - | 27.12 | 3.6 | 723120 | (24) |
| erlotinib† | 210000 | - | - | - | - | - | - | 102960 | (25) |

*see supplementary methods 1; †bioavailability estimated at 60%, ka estimated at 21.36 day$^{-1}$ (25,26)

Table 2 Pharmacodynamic Parameters

| | Value | Stoch. Approx. | |
|---|---|---|---|
| | | S.E | R.S.E.(%) |
| *Fixed Effects* | | | |
| alpha_pop | 0.17 | 0.0045 | 2.65 |
| beta_pop | 0.024 | 0.00047 | 1.95 |
| tau_pop | 0.37 | 0.013 | 3.39 |
| kk_pop | 0.0069 | 0.0013 | 19.4 |
| w_cis_rel_pop | 0.52 | 0.059 | 11.3 |
| w_car_pop | 0.82 | 0.054 | 6.53 |
| w_pem_pop | 0.87 | 0.063 | 7.32 |
| w_apo_pop | 0.78 | 0.022 | 2.87 |
| w_erl_pop | 0.66 | 0.025 | 3.77 |
| w_gem_pop | 0.95 | 0.05 | 5.27 |
| w_pac_rel_pop | 0.85 | 0.043 | 5.09 |
| w_doc_pop | 0.93 | 0.019 | 2 |
| w_bev_pop | 1.78 | 0.29 | 16.6 |
| lambda_cis_rel_pop | 0.001 | 0.000023 | 2.23 |
| lambda_car_pop | 0.0013 | 0.000092 | 7.14 |
| lambda_pem_pop | 0.0009 | 0.000039 | 4.37 |
| lambda_apo_pop | 0.00072 | 0.000014 | 1.9 |
| lambda_erl_pop | 0.0012 | 0.00011 | 8.99 |
| lambda_gem_pop | 0.0014 | 0.00005 | 3.69 |
| lambda_pac_rel_pop | 0.00088 | 0.000048 | 5.49 |
| lambda_doc_pop | 0.0011 | 0.000051 | 4.74 |
| lambda_bev_pop | 0.0039 | 0.00068 | 17.2 |
| w_bev_gamma_pop | 1.03 | 0.051 | 4.94 |
| w_bev_rho_pop | 2.05 | 0.26 | 12.7 |
| kk_i_pop | 0.0099 | 0.0025 | 25.2 |
| imv_r_perc_pop | 0.3 | 0.058 | 19.5 |
| *Standard Deviation of the Random Effects* | | | |
| omega_alpha | 0.32 | 0.02 | 6.38 |
| omega_beta | 0.23 | 0.016 | 7.11 |
| omega_tau | 0.35 | 0.033 | 9.48 |
| omega_kk | 2.06 | 0.16 | 7.55 |
| omega_w_cis_rel | 1.12 | 0.11 | 9.87 |
| omega_w_car | 0.61 | 0.059 | 9.65 |
| omega_w_pem | 0.7 | 0.061 | 8.79 |
| omega_w_apo | 0.27 | 0.026 | 9.47 |
| omega_w_erl | 0.36 | 0.32 | 8.96 |
| omega_w_gem | 0.51 | 0.042 | 8.29 |
| omega_w_pac_rel | 0.47 | 0.055 | 11.8 |
| omega_w_doc | 0.2 | 0.016 | 7.89 |
| omega_w_bev | 1.95 | 0.14 | 7.32 |
| omega_lambda_cis_rel | 0.22 | 0.02 | 8.99 |
| omega_lambda_car | 0.66 | 0.054 | 8.21 |
| omega_lambda_pem | 0.41 | 0.032 | 7.83 |
| omega_lambda_apo | 0.2 | 0.018 | 9.33 |
| omega_lambda_erl | 0.9 | 0.087 | 9.66 |
| omega_lambda_gem | 0.37 | 0.035 | 9.51 |
| *omega_lambda_pack_rel* | 0.49 | 0.05 | 10.2 |
| *omega_lambda_doc* | 0.48 | 0.043 | 9.07 |
| *omega_lambda_bev* | 1.8 | 0.12 | 6.6 |
| *omega_w_bev_gamma* | 0.47 | 0.044 | 9.36 |
| *omega_w_bev_rho* | 1.17 | 0.12 | 9.98 |
| *omega_kk_i* | 3.1 | 0.26 | 8.29 |
| *omega_imv_r_perc* | 2.85 | 0.24 | 8.52 |
| *Error Model Parameters* | | | |
| *a* | 0.11 | 0.008 | 7.57 |
| *b* | 0.088 | 0.0054 | 6.13 |

Table 3 Summary of Simulation Outcomes 1. In this simulated experiment, all patients fit during the course of the study were simulated again, except this time bevacizumab was administered between 5 and 0 days before the primary medication (m5 through c0) or between 0 and 5 days after the primary medication (c0 through p5). Below are the minimum, 5th percentile, median, 95th percentile, and maximum of the minimum tumor volume relative to baseline, e.g. in the c0 group, the median tumor reduction was by 69%.

| | *min* | *P05* | *Median* | *P95* | *Max* |
|---|---|---|---|---|---|
| *m5* | 0 | 0.00081 | 0.3 | 0.88 | 1.12 |
| *m4* | 0 | 0.00014 | 0.3 | 0.88 | 1.12 |
| *m3* | 0 | 0.000028 | 0.29 | 0.88 | 1.12 |
| *m2* | 0 | 0.00084 | 0.29 | 0.88 | 1.16 |
| *m1* | 0 | 0.0012 | 0.3 | 0.88 | 1.21 |
| *c0* | 0 | 0 | 0.31 | 0.88 | 1.25 |
| *p1* | 0 | 0.0011 | 0.31 | 0.88 | 1.3 |
| *p2* | 0 | 0.0017 | 0.31 | 0.88 | 0.135 |
| *p3* | 0 | 0.0017 | 0.3 | 0.89 | 1.39 |
| *p4* | 0 | 0.0019 | 0.3 | 0.89 | 1.44 |
| *p5* | 0 | 0.0015 | 0.29 | 0.89 | 1.49 |

Table 4 Summary of Simulation Outcomes 2. In this simulated experiment, bevacizumab pemetrexed and cisplatin were administered at recommended dosages to virtual patients every 21 days for 4 cycles. The gap between bevacizumab and pemetrexed/cisplatin administration was set at either 5 days (m5), 3 days (m3), 1 day (m1), 0 days (c0), or pemetrexed/cisplatin was administered either 1 day (p1) or 2 days (p2) before bevacizumab. Below are the minimum, 5th percentile, median, 95th percentile, and maximum of the minimum tumor volume relative to baseline, e.g. in the c0 group, the median tumor reduction was by 39%

|     | min | P05  | Median | P95 | Max  |
|-----|-----|------|--------|-----|------|
| m5  | 0   | 0.11 | 0.65   | 1   | 1.07 |
| m3  | 0   | 0.13 | 0.63   | 1   | 1.07 |
| m1  | 0   | 0.15 | 0.61   | 1   | 1.09 |
| c0  | 0   | 0.16 | 0.6    | 1   | 1.09 |
| p1  | 0   | 0.16 | 0.59   | 1   | 1.09 |
| p2  | 0   | 0.16 | 0.59   | 1   | 1.09 |

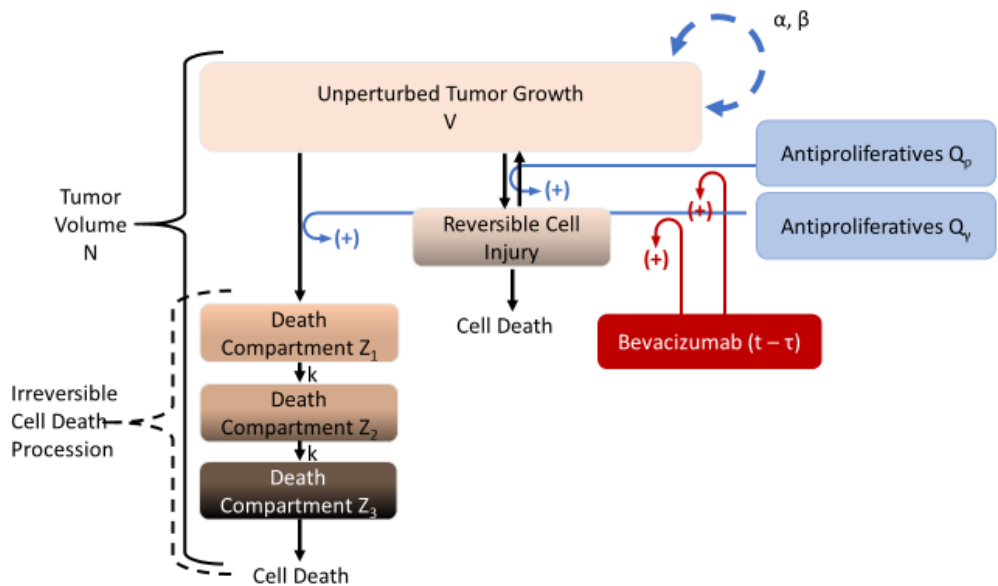

Figure 1 Model Diagram.

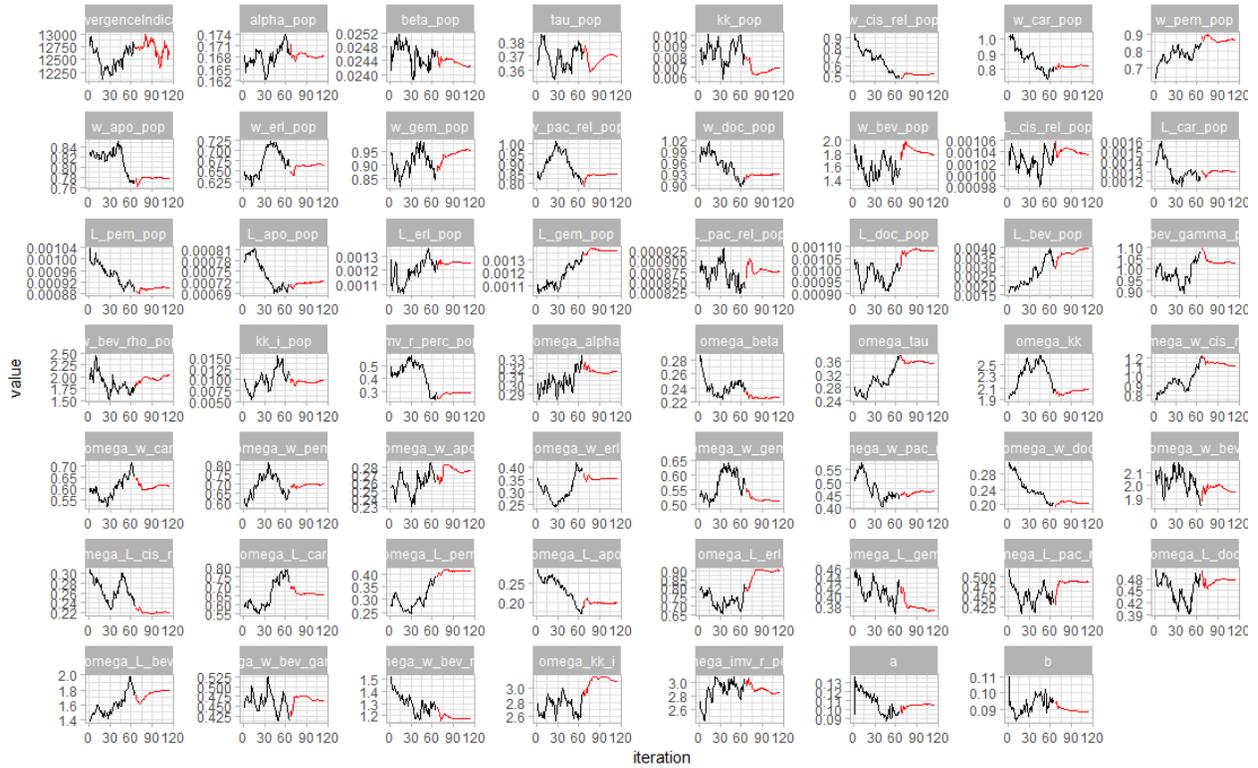

Figure 2 SAEM Search. Stochastic approximation expectation maximization search for most likely estimates of parameter values. Exploratory search in black and smoothing search in red.

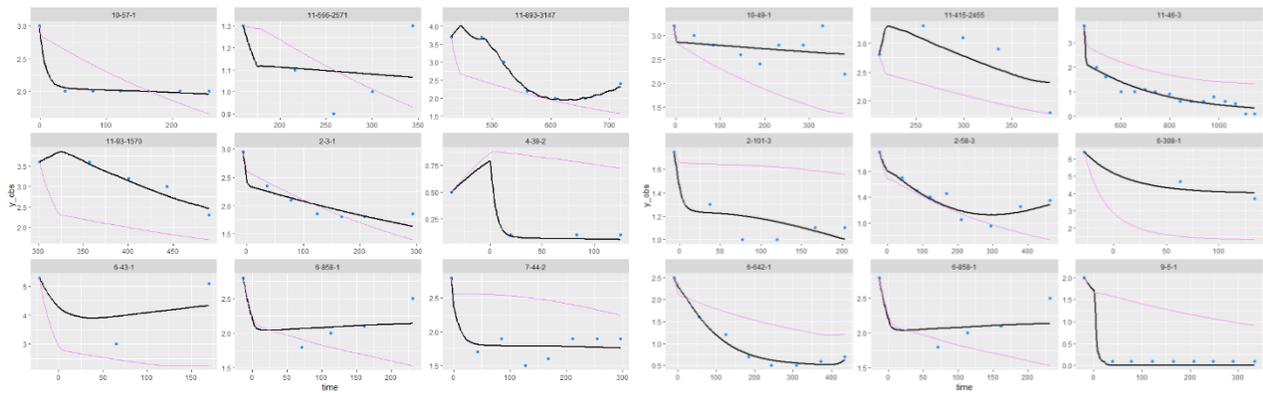

Figure 3 Sample of Individual Fits. Several individual fits (black) vs observations (blue) and population prediction (purple).

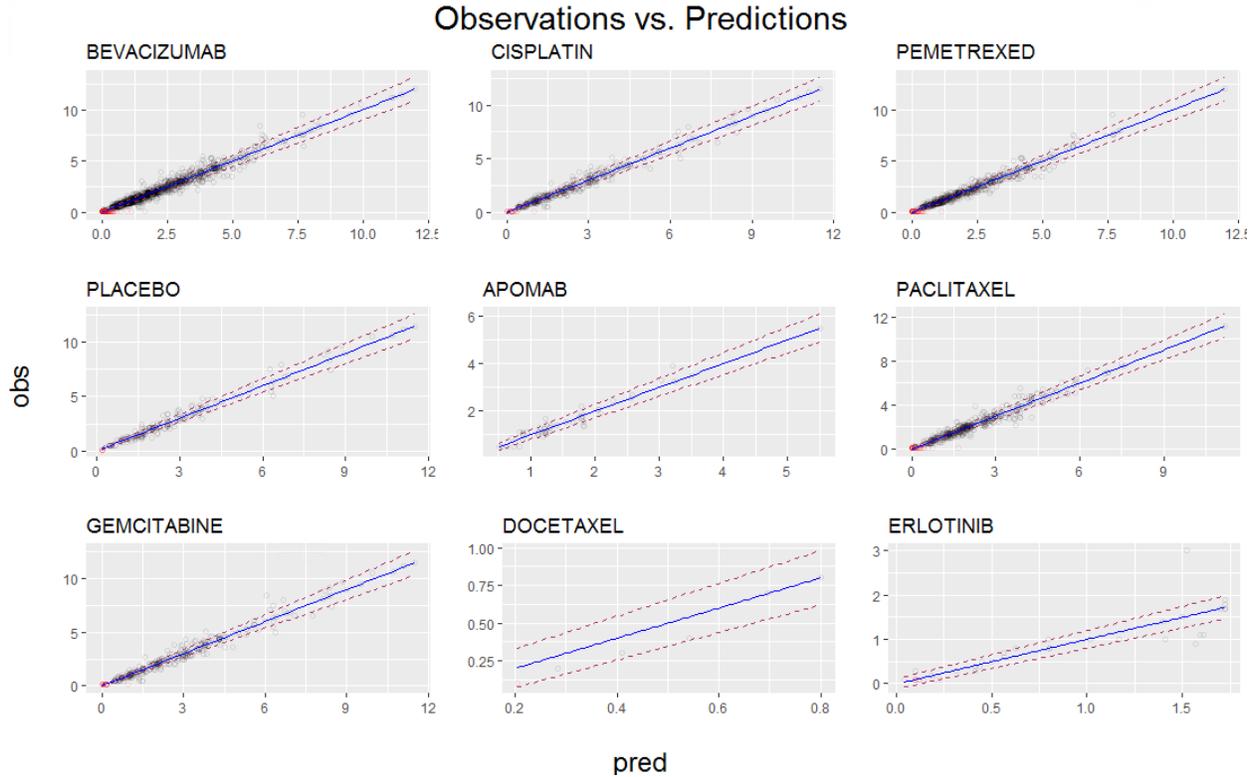

Figure 4 Observations vs. Predictions by Patients Receiving Therapeutic. In these figures the observations vs predictions (black points) are plotted along with censored data (red points). Blue line is where observations meet predictions i.e. ratio is 1. Error model 95% prediction boundaries at dotted red lines. Points are semitransparent to reduce visual overcrowding of points.

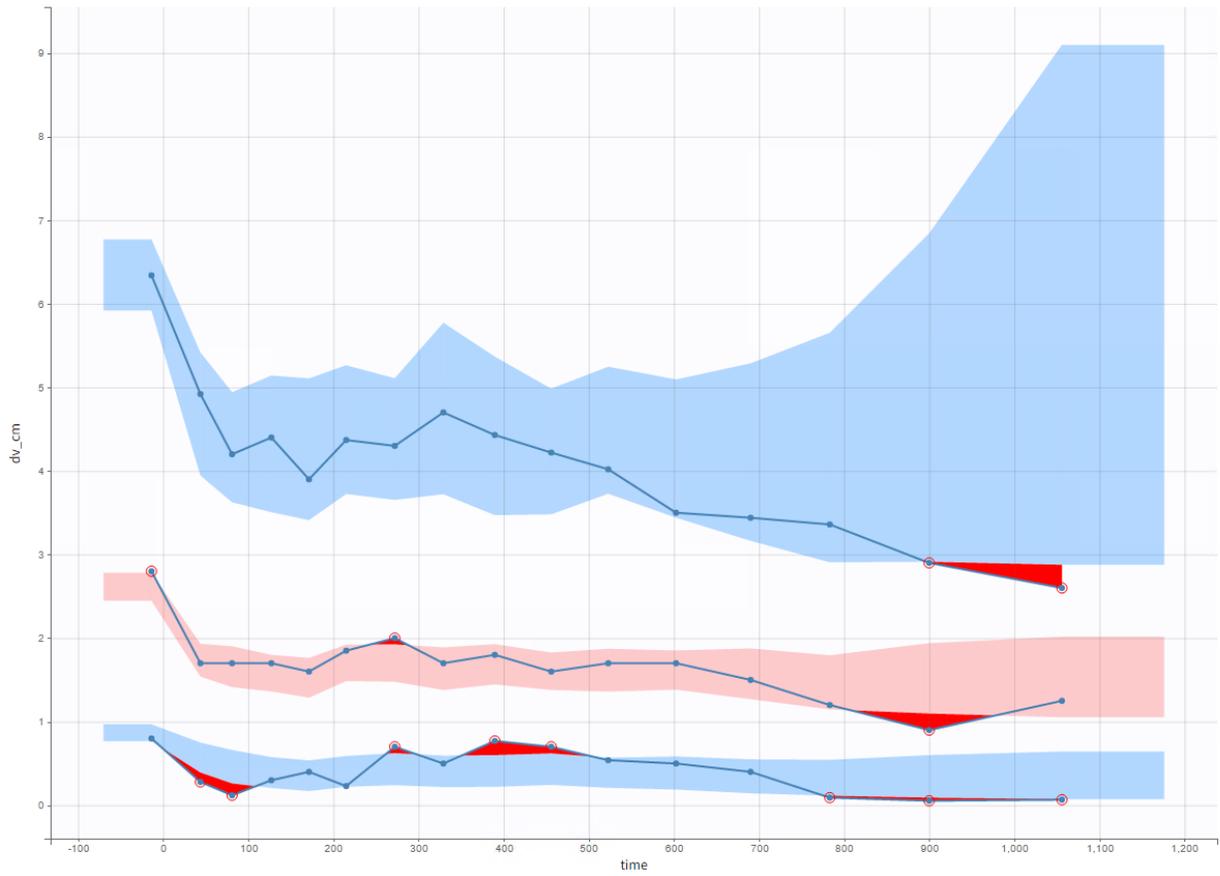

Figure 5 Visual Predictive Check.

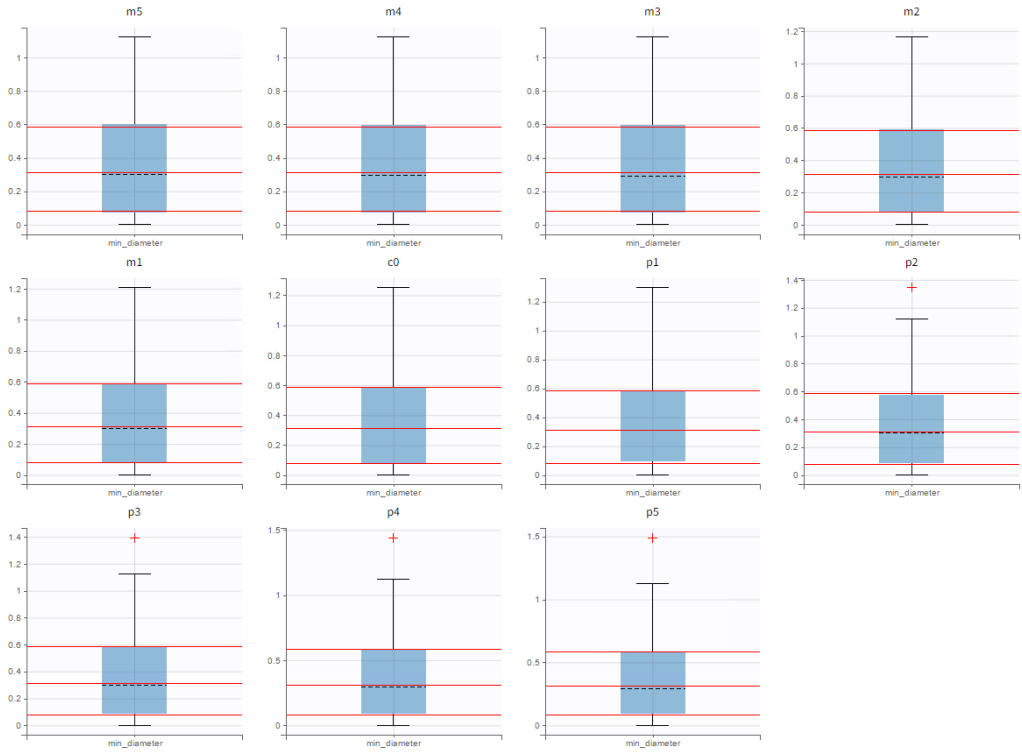

Figure 6 Summary of Simulation Outcomes with Box and Wiskers Plots 1. In this simulated experiment, all patients fit during the course of the study were simulated again, except this time bevacizumab was administered between 5 and 0 days before the primary medication (m5 through c0) or between 0 and 5 days after the primary medication (c0 through p5). The horizontal red lines are reference lines to the 1st, 2nd, and 3rd quartile for the simulated trial where gap was equal to zero.

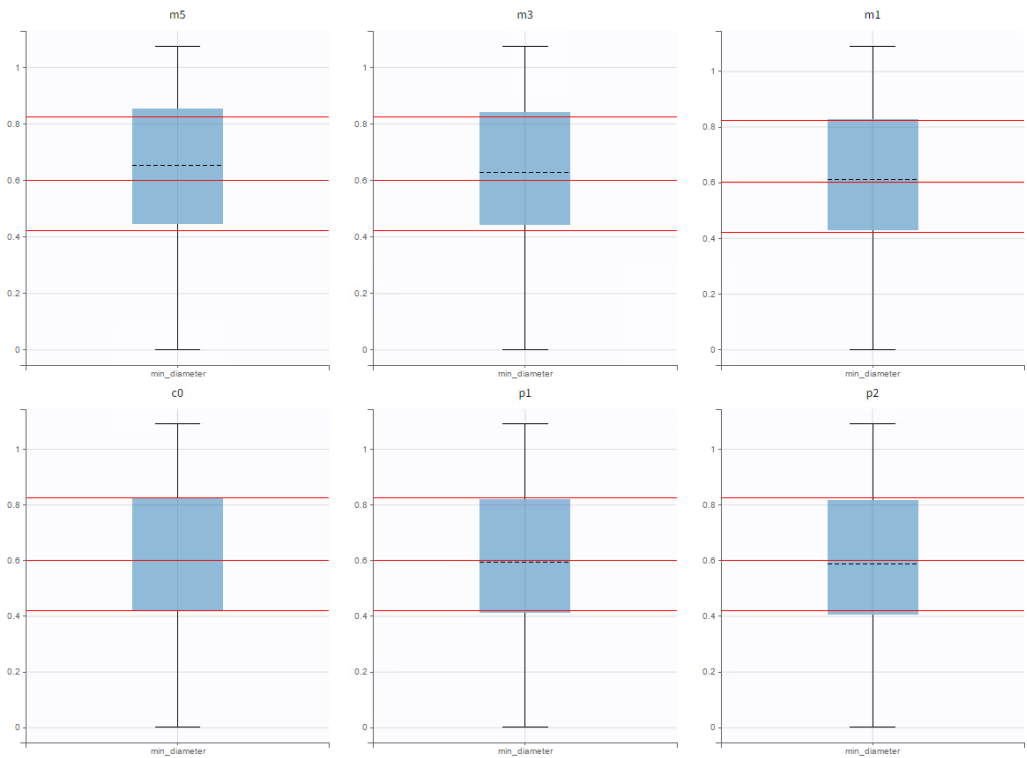

Figure 7 Summary of Simulation Outcomes with Box and Wiskers Plots 2. In this simulated experiment, bevacizumab pemetrexed and cisplatin were administered at recommended dosages to virtual patients every 21 days for 4 cycles. The gap between bevacizumab and pemetrexed/cisplatin administration was set at either 5 days (m5), 3 days (m3), 1 day (m1), 0 days (c0), or pemetrexed/cisplatin was administered either 1 day (p1) or 2 days (p2) before bevacizumab. The horizontal red lines are reference lines to the 1st, 2nd, and 3rd quartile for the simulated trial where gap was equal to zero.

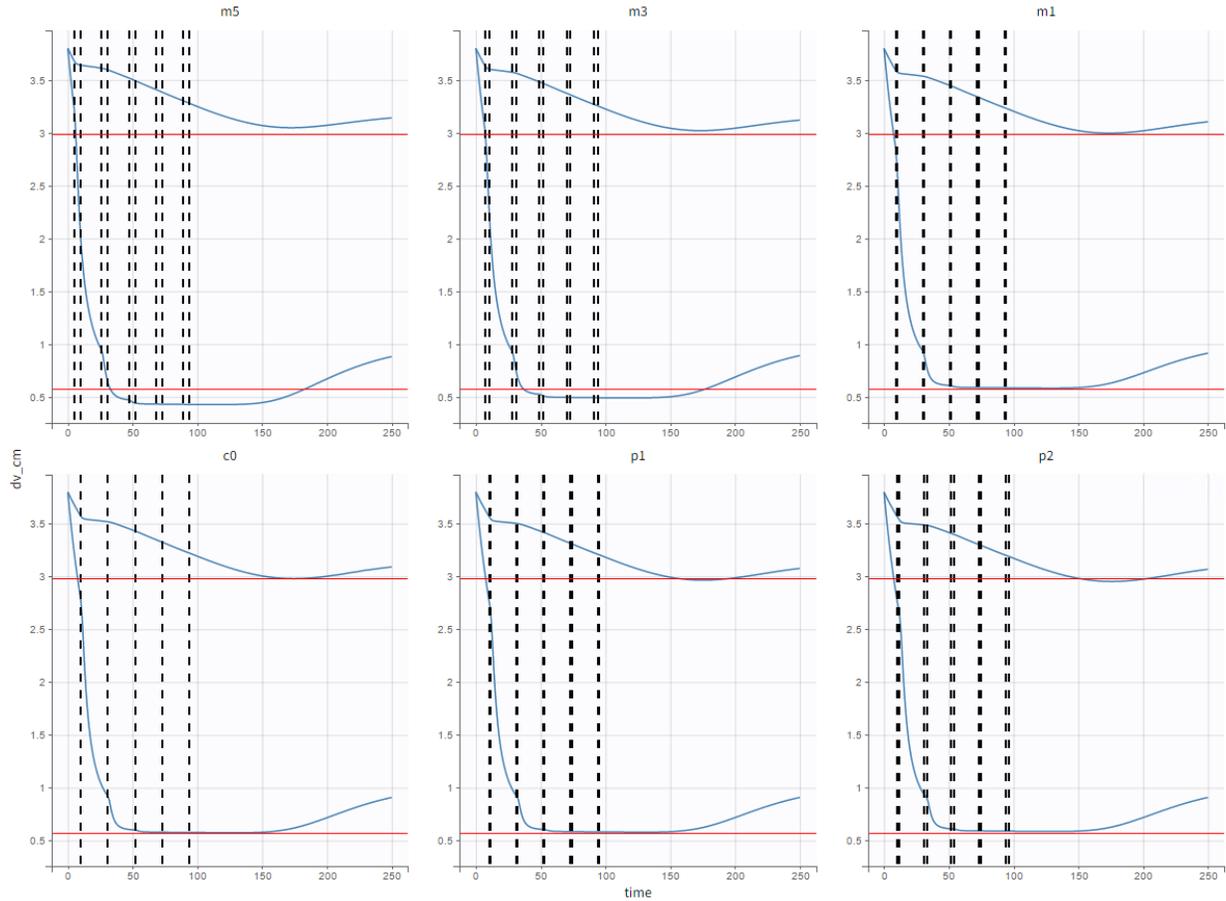

Figure 8 A Pair of Illustrative Examples. In this simulated experiment, bevacizumab pemetrexed and cisplatin were administered at recommended dosages to virtual patients every 21 days for 4 cycles. The gap between bevacizumab and pemetrexed/cisplatin administration was set at either 5 days (m5), 3 days (m3), 1 day (m1), 0 days (c0), or pemetrexed/cisplatin was administered either 1 day (p1) or 2 days (p2) before bevacizumab. For the virtual patient with a larger final tumor volume, administering bevacizumab after pemetrexed/cisplatin produces the maximum tumor volume reduction. However, the opposite is true of the virtual patient with the greater response.

# References

[1] Global Cancer Observatory: Cancer Today. Lyon, France: International Agency for Research on Cancer. [cited 2020 Nov 20]. Available from: http://gco.iarc.fr/today/home

[2] Chang A. Chemotherapy, chemoresistance and the changing treatment landscape for NSCLC. Lung Cancer. 2011 Jan;71(1):3–10.

[3] Siegel RL, Miller KD, Jemal A. Cancer statistics, 2020. CA Cancer J Clin. 2020;70(1):7–30.

[4] Barlesi F, Imbs D-C, Tomasini P, Greillier L, Galloux M, Testot-Ferry A, et al. Mathematical modeling for Phase I cancer trials: A study of metronomic vinorelbine for advanced non-small cell lung cancer (NSCLC) and mesothelioma patients. Oncotarget. 2017 May 2;8(29):47161–6.

[5] Simeoni M, Magni P, Cammia C, Nicolao GD, Croci V, Pesenti E, et al. Predictive Pharmacokinetic-Pharmacodynamic Modeling of Tumor Growth Kinetics in Xenograft Models after Administration of Anticancer Agents. Cancer Res. 2004 Feb 1;64(3):1094–101.

[6] Benzekry S, Pasquier E, Barbolosi D, Lacarelle B, Barlési F, André N, et al. Metronomic reloaded: Theoretical models bringing chemotherapy into the era of precision medicine. Semin Cancer Biol. 2015 Dec 1;35:53–61.

[7] Agur Z, Elishmereni M, Foryś U, Kogan Y. Accelerating the Development of Personalized Cancer Immunotherapy by Integrating Molecular Patients' Profiles with Dynamic Mathematical Models. Clin Pharmacol Ther. 2020;108(3):515–27.

[8] Ferrara N, Hillan KJ, Gerber H-P, Novotny W. Discovery and development of bevacizumab, an anti-VEGF antibody for treating cancer. Nat Rev Drug Discov. 2004 May;3(5):391–400.

[9] Schneider BK, Boyer A, Ciccolini J, Barlesi F, Wang K, Benzekry S, et al. Optimal Scheduling of Bevacizumab and Pemetrexed/Cisplatin Dosing in Non-Small Cell Lung Cancer. CPT Pharmacomet Syst Pharmacol. 2019;8(8):577–86.

[10] Schneider B, Balbas-Martinez V, Jergens AE, Troconiz IF, Allenspach K, Mochel JP. Model-Based Reverse Translation Between Veterinary and Human Medicine: The One Health Initiative: Model-Based Reverse Translational Pharmacology. CPT Pharmacomet Syst Pharmacol. 2018 Feb;7(2):65–8.

[11] Musser ML, Mahaffey AL, Fath MA, Buettner GR, Wagner BA, Schneider BK, et al. In vitro Cytotoxicity and Pharmacokinetic Evaluation of Pharmacological Ascorbate in Dogs. Front Vet Sci [Internet]. 2019 [cited 2020 May 11];6. Available from: https://www.frontiersin.org/articles/10.3389/fvets.2019.00385/full

[12] Imbs D-C, Cheikh RE, Boyer A, Ciccolini J, Mascaux C, Lacarelle B, et al. Revisiting Bevacizumab + Cytotoxics Scheduling Using Mathematical Modeling: Proof of Concept Study in Experimental Non-Small Cell Lung Carcinoma. CPT Pharmacomet Syst Pharmacol. 2018;7(1):42–50.

[13] Model-Based Prediction of Phase III Overall Survival in Colorectal Cancer on the Basis of Phase II Tumor Dynamics | Journal of Clinical Oncology [Internet]. [cited 2020 Mar 23]. Available from: https://ascopubs.org/doi/pdf/10.1200/JCO.2008.21.0807

[14] Spratt JS, Meyer JS, Spratt JA. Rates of growth of human solid neoplasms: Part I. J Surg Oncol. 1995 Oct;60(2):137–46.

[15] Defining a data set for Monolix [Internet]. Monolix 2017. [cited 2020 Nov 21]. Available from: http://monolix.lixoft.com/data-and-models/creating-data-set/

[16] Common Terminology Criteria for Adverse Events (CTCAE). 2017;155.

[17] Lu J-F, Bruno R, Eppler S, Novotny W, Lum B, Gaudreault J. Clinical pharmacokinetics of bevacizumab in patients with solid tumors. Cancer Chemother Pharmacol. 2008 Oct;62(5):779–86.

[18] Urien S, Brain E, Bugat R, Pivot X, Lochon I, Van M-LV, et al. Pharmacokinetics of platinum after oral or intravenous cisplatin: a phase 1 study in 32 adult patients. Cancer Chemother Pharmacol. 2005 Jan;55(1):55–60.

[19] Latz JE, Chaudhary A, Ghosh A, Johnson RD. Population pharmacokinetic analysis of ten phase II clinical trials of pemetrexed in cancer patients. Cancer Chemother Pharmacol. 2006 Apr;57(4):401–11.

[20] Camidge DR, Herbst RS, Gordon MS, Eckhardt SG, Kurzrock R, Durbin B, et al. A Phase I Safety and Pharmacokinetic Study of the Death Receptor 5 Agonistic Antibody PRO95780 in Patients with Advanced Malignancies. Clin Cancer Res. 2010 Feb 15;16(4):1256–63.

[21] Henningsson A, Karlsson MO, Viganò L, Gianni L, Verweij J, Sparreboom A. Mechanism-Based Pharmacokinetic Model for Paclitaxel. J Clin Oncol. 2001 Oct 15;19(20):4065–73.

[22] Joerger M, Huitema ADR, Richel DJ, Dittrich C, Pavlidis N, Briasoulis E, et al. Population Pharmacokinetics and Pharmacodynamics of Paclitaxel and Carboplatin in Ovarian Cancer Patients: A Study by the European Organization for Research and Treatment of Cancer-Pharmacology and Molecular Mechanisms Group and New Drug Development Group. Clin Cancer Res. 2007 Nov 1;13(21):6410–8.

[23] Jiang X, Galettis P, Links M, Mitchell PL, McLachlan AJ. Population pharmacokinetics of gemcitabine and its metabolite in patients with cancer: effect of oxaliplatin and infusion rate. Br J Clin Pharmacol. 2008 Mar;65(3):326–33.

[24] Slaviero KA, Clarke SJ, McLachlan AJ, Blair EYL, Rivory LP. Population pharmacokinetics of weekly docetaxel in patients with advanced cancer. Br J Clin Pharmacol. 2004 Jan;57(1):44–53.

[25] Lu J-F, Eppler SM, Wolf J, Hamilton M, Rakhit A, Bruno R, et al. Clinical pharmacokinetics of erlotinib in patients with solid tumors and exposure-safety relationship in patients with non–small cell lung cancer. Clin Pharmacol Ther. 2006;80(2):136–45.

[26] Frohna P, Lu J, Eppler S, Hamilton M, Wolf J, Rakhit A, et al. Evaluation of the absolute oral bioavailability and bioequivalence of erlotinib, an inhibitor of the epidermal growth factor receptor tyrosine kinase, in a randomized, crossover study in healthy subjects. J Clin Pharmacol. 2006 Mar;46(3):282–90.

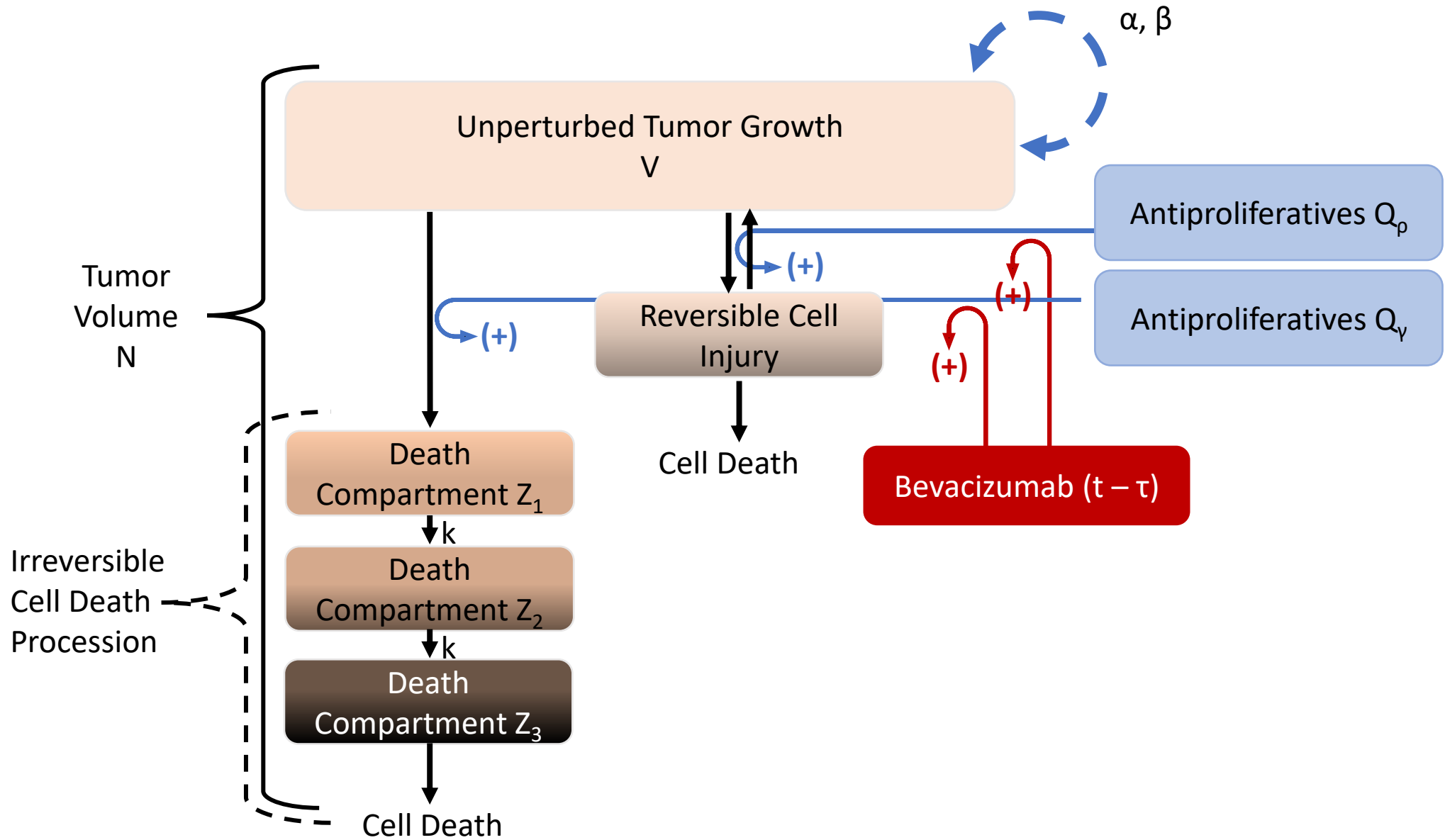

**Figure 1** Model Diagram.

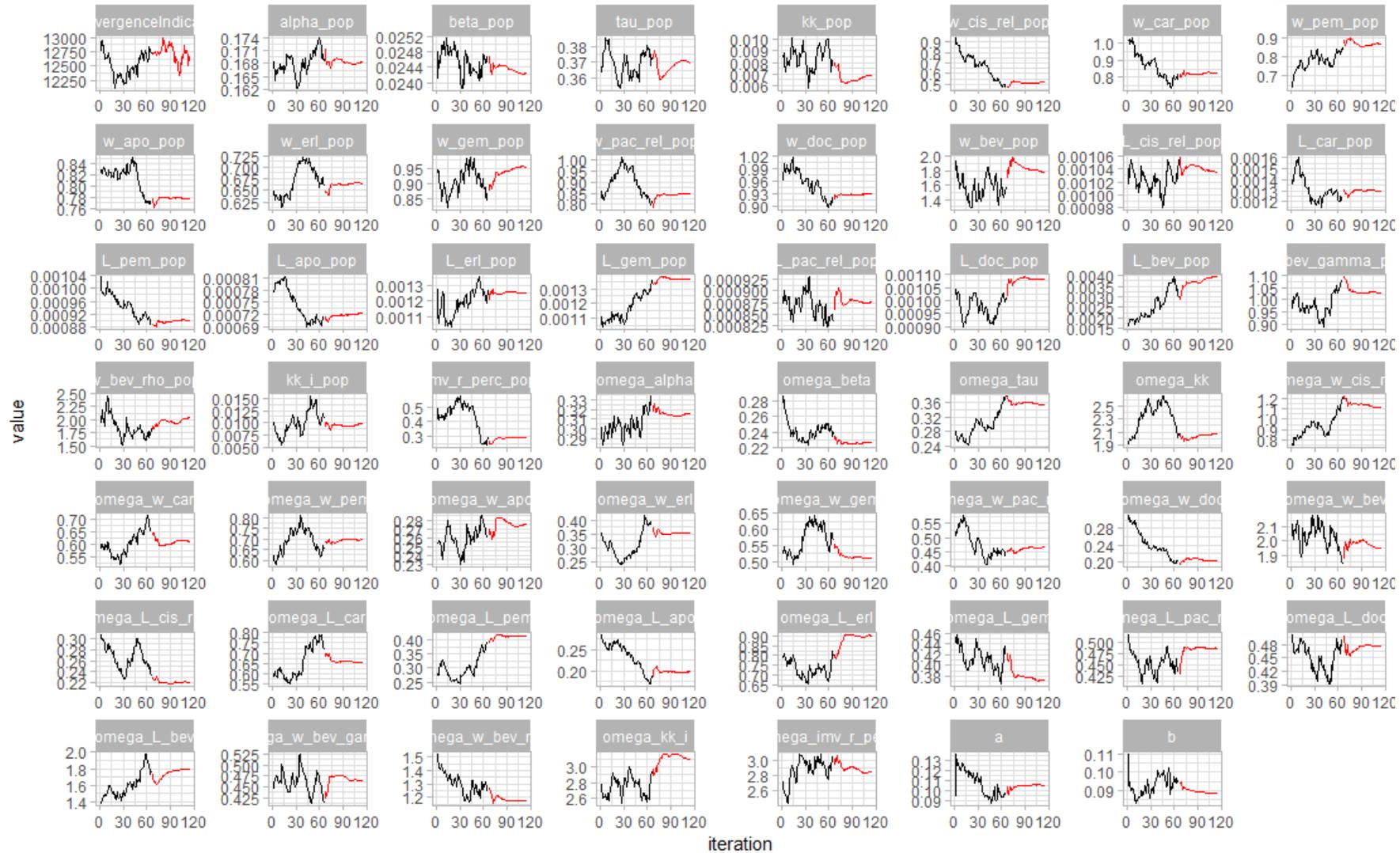

**Figure 2** SAEM Search. Stochastic approximation expectation maximization search for most likely estimates of parameter values. Exploratory search in black and smoothing search in red.

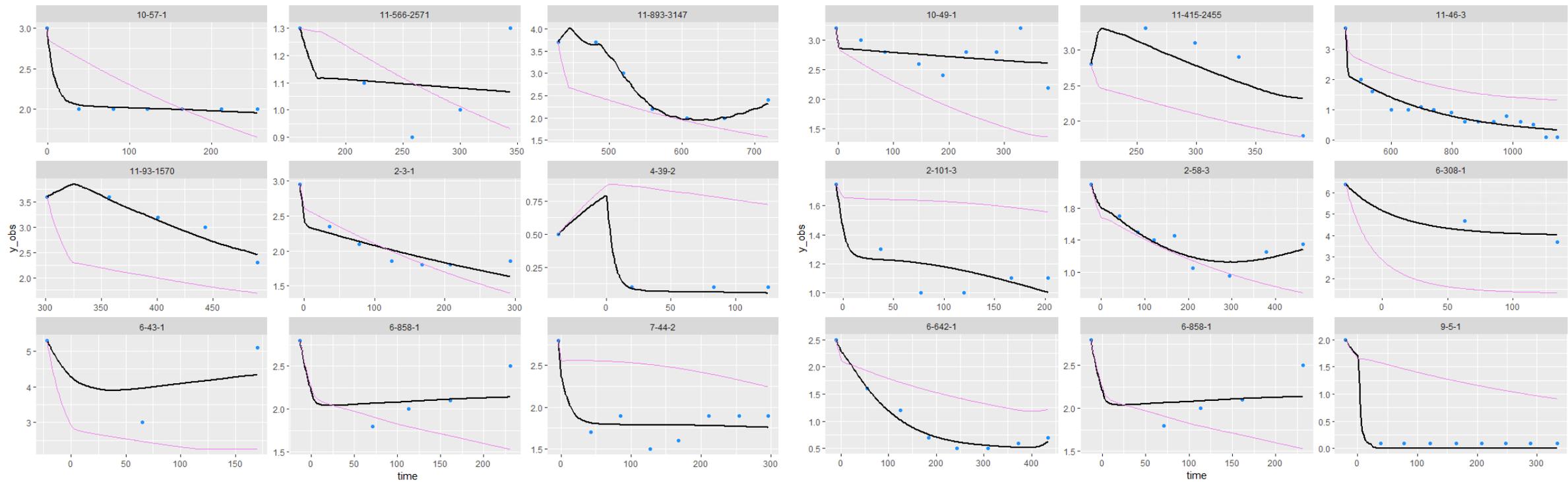

**Figure 3** Sample of Individual Fits. Several individual fits (black) vs observations (blue) and population prediction (purple).

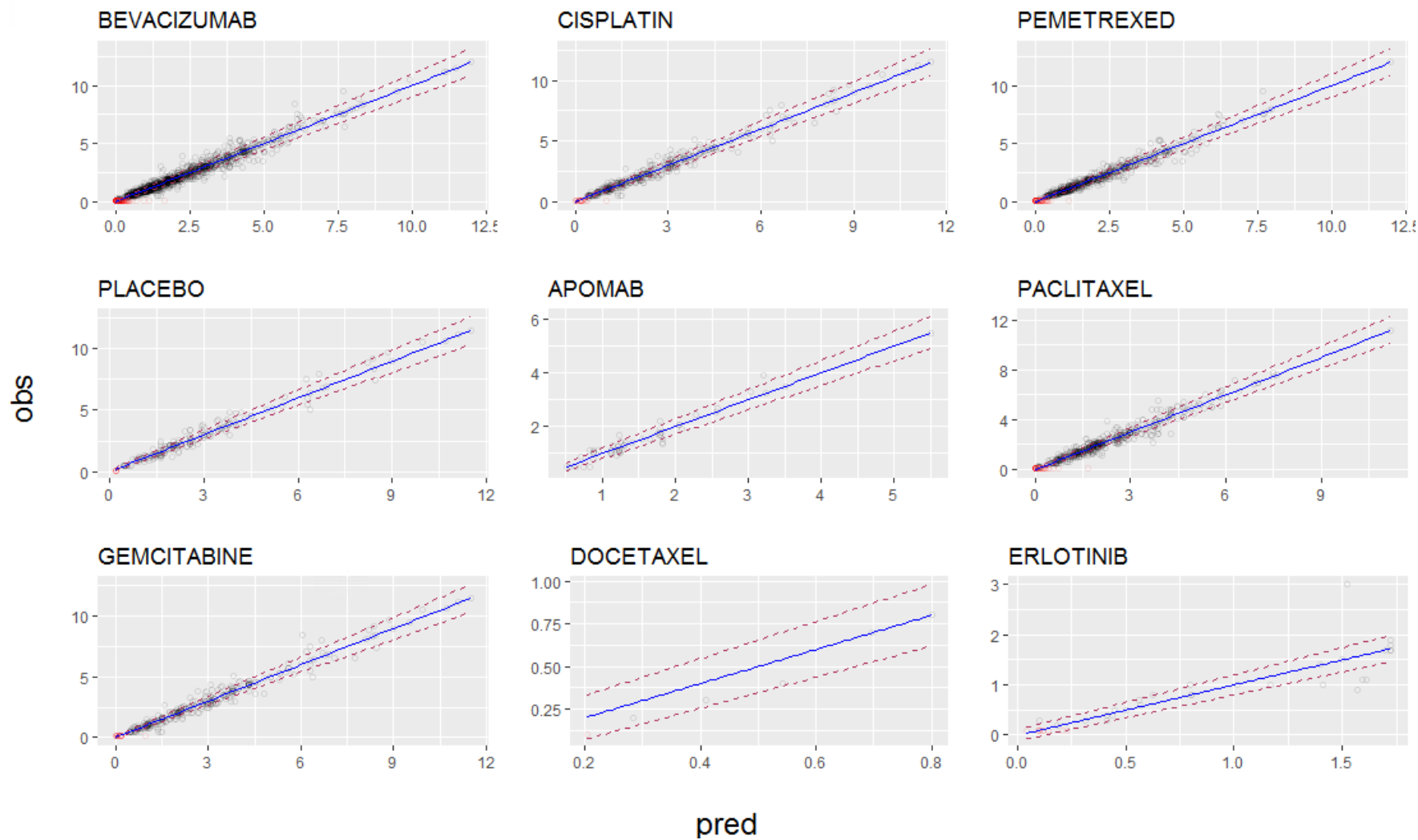

**Figure 4** Observations vs. Predictions by Patients Receiving Therapeutic. In these figures the observations vs predictions (black points) are plotted along with censored data (red points). Blue line is where observations meet predictions i.e. ratio is 1. Error model 95% prediction boundaries at dotted red lines. Points are semitransparent to reduce visual overcrowding of points.

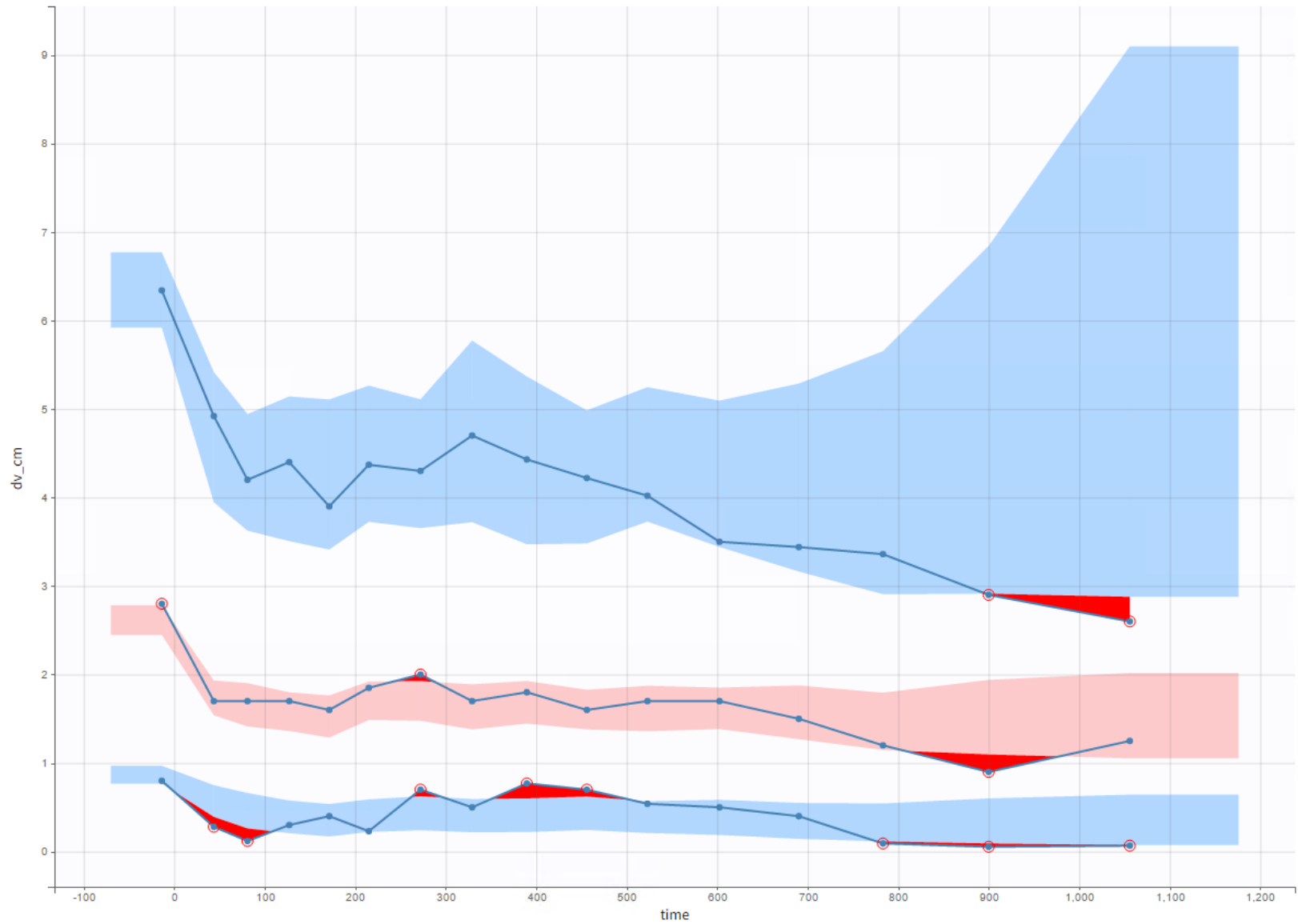

**Figure 5** Visual Predictive Check.

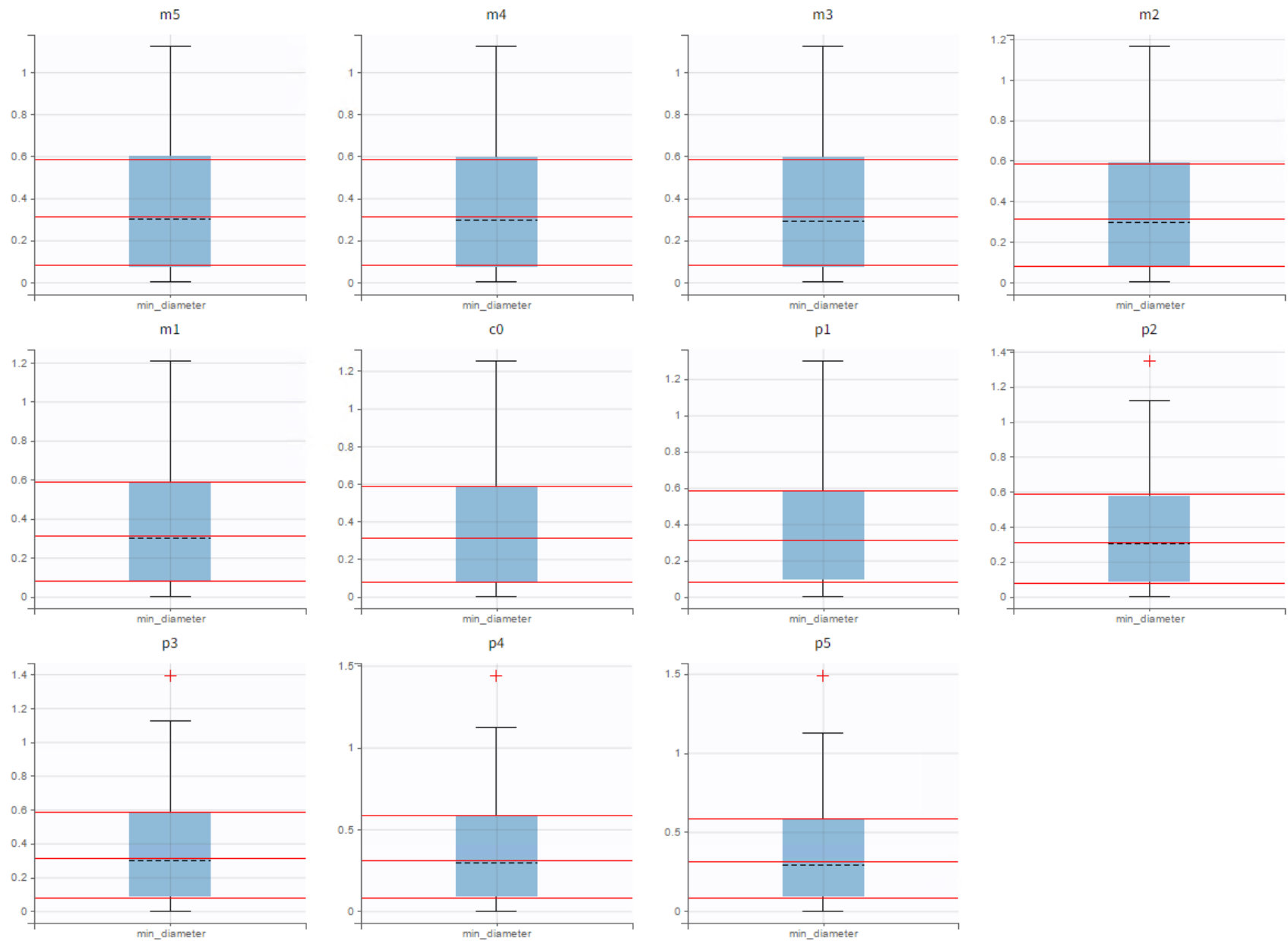

**Figure 6** Summary of Simulation Outcomes with Box and Wiskers Plots 1. In this simulated experiment, all patients fit during the course of the study were simulated again, except this time bevacizumab was administered between 5 and 0 days before the primary medication (m5 through c0) or between 0 and 5 days after the primary medication (c0 through p5). The horizontal red lines are reference lines to the 1st, 2nd, and 3rd quartile for the simulated trial where gap was equal to zero.

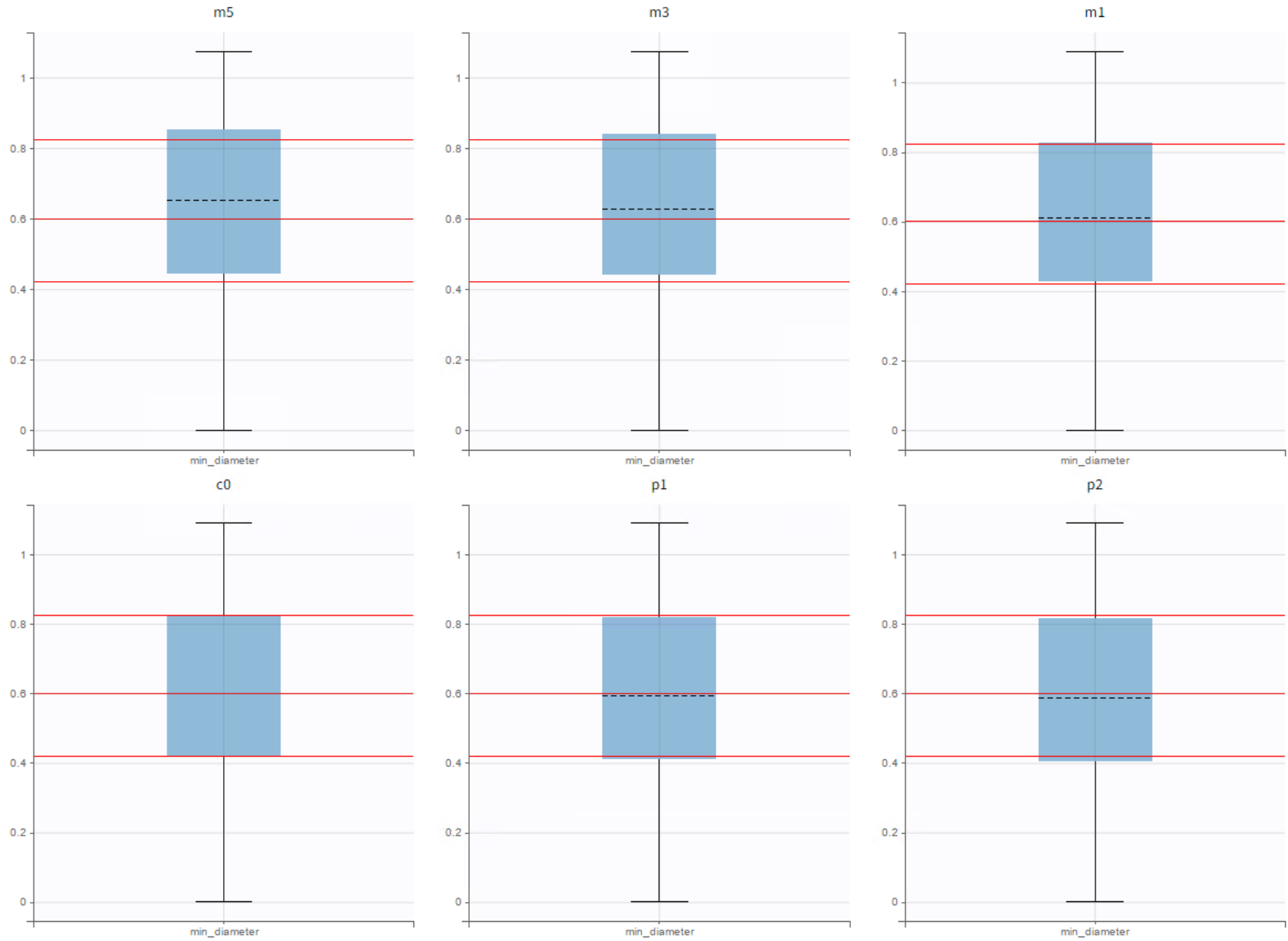

**Figure 7** Summary of Simulation Outcomes with Box and Wiskers Plots 2. In this simulated experiment, bevacizumab pemetrexed and cisplatin were administered at recommended dosages to virtual patients every 21 days for 4 cycles. The gap between bevacizumab and pemetrexed/cisplatin administration was set at either 5 days (m5), 3 days (m3), 1 day (m1), 0 days (c0), or pemetrexed/cisplatin was administered either 1 day (p1) or 2 days (p2) before bevacizumab. The horizontal red lines are reference lines to the 1st, 2nd, and 3rd quartile for the simulated trial where gap was equal to zero.

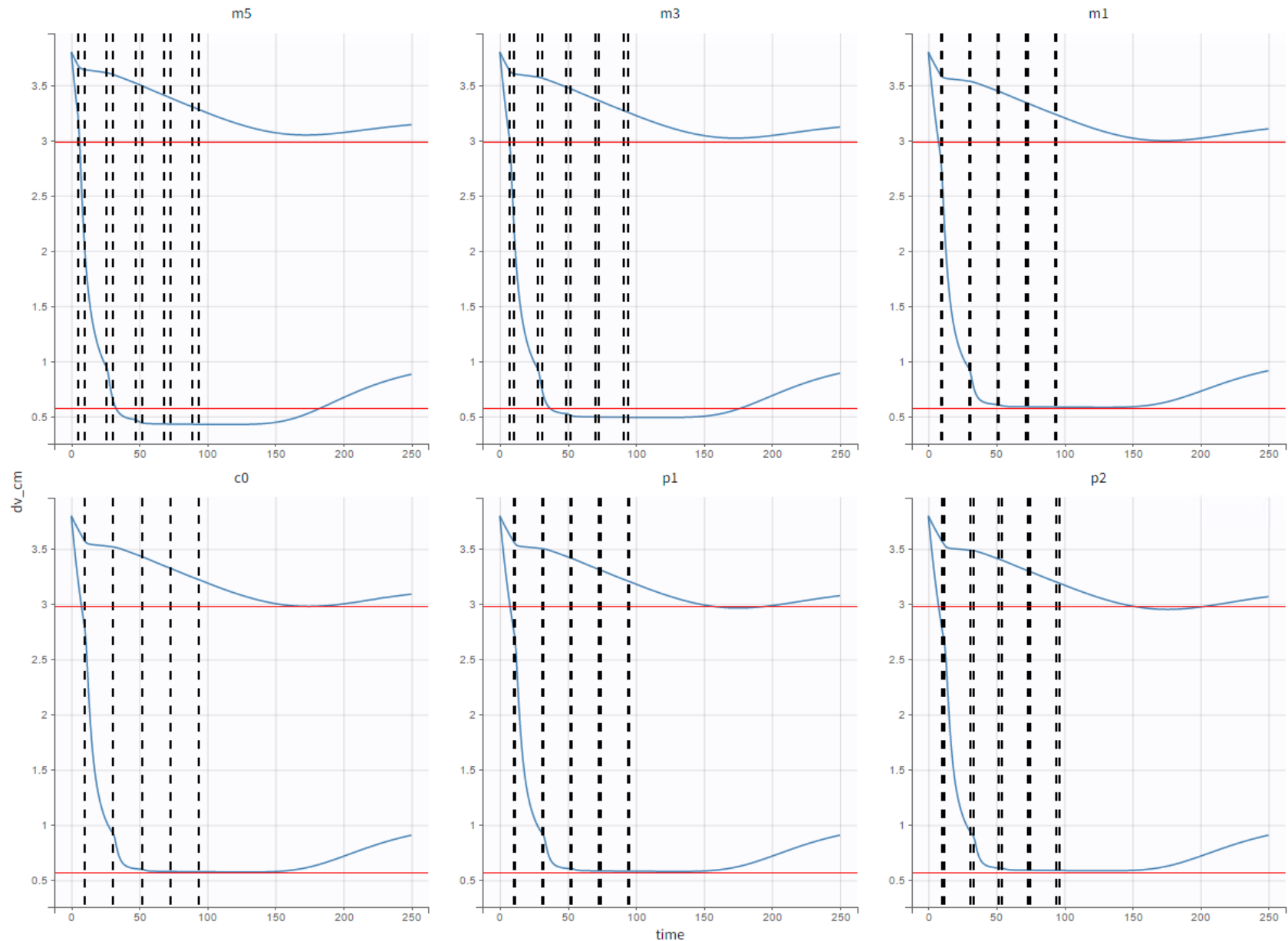

**Figure 8** A Pair of Illustrative Examples. In this simulated experiment, bevacizumab pemetrexed and cisplatin were administered at recommended dosages to virtual patients every 21 days for 4 cycles. The gap between bevacizumab and pemetrexed/cisplatin administration was set at either 5 days (m5), 3 days (m3), 1 day (m1), 0 days (c0), or pemetrexed/cisplatin was administered either 1 day (p1) or 2 days (p2) before bevacizumab. For the virtual patient with a larger final tumor volume, administering bevacizumab after pemetrexed/cisplatin produces the maximum tumor volume reduction. However, the opposite is true of the virtual patient with the greater response.